\documentclass[aps,prl,twocolumn,showpacs,10pt,superscriptaddress,preprintnumbers,nofootinbib]{revtex4-2}
\usepackage{amsmath,bm}
\usepackage{physics}
\usepackage{graphicx, xcolor}
\usepackage{amssymb}
\usepackage[colorlinks, linkcolor=blue!80!black, citecolor=green!50!black, urlcolor=magenta]{hyperref}
\usepackage[normalem]{ulem}

\newcommand{\beq}[1][]{\begin{equation}\label{#1}}
\newcommand{\eeq}{\end{equation}}
\newcommand{\bse}{\begin{subequations}}
\newcommand{\ese}{\end{subequations}}
\newcommand{\nn}{\nonumber}

\renewcommand{\pb}{{\rm pb}}

\newcommand{\ab}{{\rm ab}}

\newcommand{\eq}[1]{Eq.~\eqref{#1}}
\newcommand{\fig}[1]{Fig.~\ref{#1}}

\newcommand{\pp}[1]{\left( #1 \right)}
\newcommand{\bb}[1]{\left[ #1 \right]}

\newcommand{\htt}{ht\bar{t}}

\renewcommand{\k}{\kappa}

\newcommand{\kt}{\widetilde{\kappa}}
\newcommand{\lum}{\mathcal{L}}
\newcommand{\M}{\mathcal{M}}
\newcommand{\A}{\mathcal{A}}
\newcommand{\At}{\widetilde{\mathcal{A}}}

\newcommand{\GeV}{{\rm GeV}}
\newcommand{\TeV}{{\rm TeV}}
\newcommand{\sgn}{{\rm sgn}}

\begin{document}

\preprint{
	{\vbox {			
		\hbox{\bf MSUHEP-22-034}
}}}
\vspace*{0.2cm}

\title{Determining the $CP$ Property of $\htt$ Coupling via a Novel Jet Substructure Observable}

\author{Zhite Yu}
\email{yuzhite@msu.edu (corresponding author)}
\affiliation{Department of Physics and Astronomy,
Michigan State University, East Lansing, Michigan 48824, USA}

\author{Kirtimaan A. Mohan}
\email{kamohan@msu.edu}
\affiliation{Department of Physics and Astronomy,
Michigan State University, East Lansing, Michigan 48824, USA}

\author{C.-P. Yuan}
\email{yuanch@msu.edu}
\affiliation{Department of Physics and Astronomy,
Michigan State University, East Lansing, Michigan 48824, USA}

\date{\today}

\begin{abstract}
Determining the $CP$ property of the Higgs boson is important for a precision test of the Standard Model as well as for the search for new physics. We propose a novel jet substructure observable based on the azimuthal anisotropy in a linearly polarized gluon jet that is produced in association with a Higgs boson at hadron colliders, and demonstrate that it provides a new $CP$-odd observable for determining the $CP$ property of the Higgs-top interaction. 
We introduce a factorization formalism to define a polarized gluon jet function with the insertion of an infrared-safe azimuthal observable to capture the linear polarization.
\end{abstract}

\maketitle


\emph{Introduction.}---Pinning down the $CP$ nature of the Higgs-top interaction ($\htt$) is an important program being pursued at the Large Hadron Collider (LHC)~\cite{CMS:2021nnc, CMS:2020cga, ATLAS:2020ior, CMS:2022dbt, ATLAS:2020ior, ATLAS:2023cbt}. 
Any deviation from a Standard-Model-like $\htt$ coupling could indicate new physics as well as provide a potential source for the $CP$ violation as required by the baryogenesis~\cite{Sakharov:1967dj}.
Unlike $CP$-violating Higgs interactions with vector bosons, which arise from dimension-six operators, $CP$-violating effects in the $\htt$ coupling could occur via a dimension-four operator, 
\beq[eq:L]
	\mathcal{L} \supset - \frac{y_t}{\sqrt{2}} \, h \, \bar{t} \, (\k + i \,\kt\, \gamma_5) \, t \,,
\eeq
and can be potentially larger. In \eq{eq:L}, $y_t = \sqrt{2} m_t / v$ is the Yukuwa coupling of Higgs and top quark in the Standard Model (SM), and $(\k, \kt)$ parametrize the $CP$-even and $CP$-odd $\htt$ interactions, respectively, which can be reparametrized as $(\k, \kt) = \k_t (\cos\alpha, \sin\alpha)$, with $\alpha$ being the $CP$ phase. The SM corresponds to $(\k, \kt) = (1, 0)$ or $(\k_t, \alpha) = (1, 0)$.

Numerous approaches have been proposed for determining the $CP$ phase, 
either directly via associated Higgs and top production~\cite{Ellis:2013yxa, Boudjema:2015nda, Buckley:2015vsa, Gritsan:2016hjl, Mileo:2016mxg, AmorDosSantos:2017ayi, Azevedo:2017qiz, Li:2017dyz, Goncalves:2018agy, Faroughy:2019ird, Bortolato:2020zcg, Cao:2020hhb, Goncalves:2021dcu, Patrick:2019nhv},
or indirectly via Higgs or top induced loop effects~\cite{Hankele:2006ja, Brod:2013cka, Dolan:2014upa, Englert:2012xt, Bernlochner:2018opw, Englert:2019xhk, Gritsan:2020pib, Bahl:2020wee, Martini:2021uey}. 
The sensitivity to $\alpha$ can be enhanced by using observables that are odd 
under $CP$ transformation~\cite{Mileo:2016mxg, Goncalves:2018agy}. 
Machine learning techniques have also been considered~\cite{Patrick:2019nhv, Ren:2019xhp, Bortolato:2020zcg, Bahl:2021dnc, Barman:2021yfh} 
to optimize the sensitivity.
The current experimental bounds from direct measurements for various Higgs detection channels are 
$\abs{\alpha} \leq 35^\circ$~\citep{CMS:2020cga}, 
$48^\circ$~\citep{CMS:2022dbt}, 
and $63^\circ$~\citep{ATLAS:2023cbt} at $68\%$ C.L.,
and 
$\abs{\alpha} \leq 43^\circ$ at $95\%$ C.L.~\citep{ATLAS:2020ior}.
These need to be further constrained by more complementary observables, 
at the upcoming High-Luminosity LHC (HL-LHC)~\citep{Apollinari:2120673} 
and a possible future $pp$ collider at $100~\TeV$ (FCC-hh)~\citep{Mangano:2270978}.

\begin{figure}[htbp]
	\centering
		\includegraphics[trim={0.45cm -0.32cm 0.3cm 0.4cm}, clip, scale=0.8]{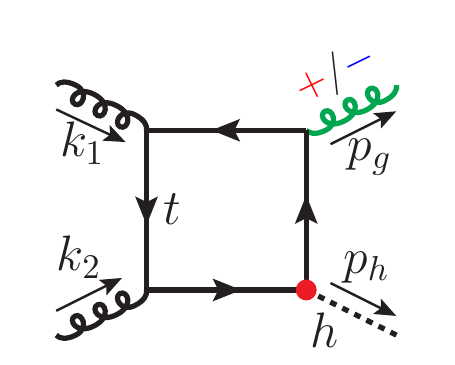}
		\includegraphics[scale=0.55]{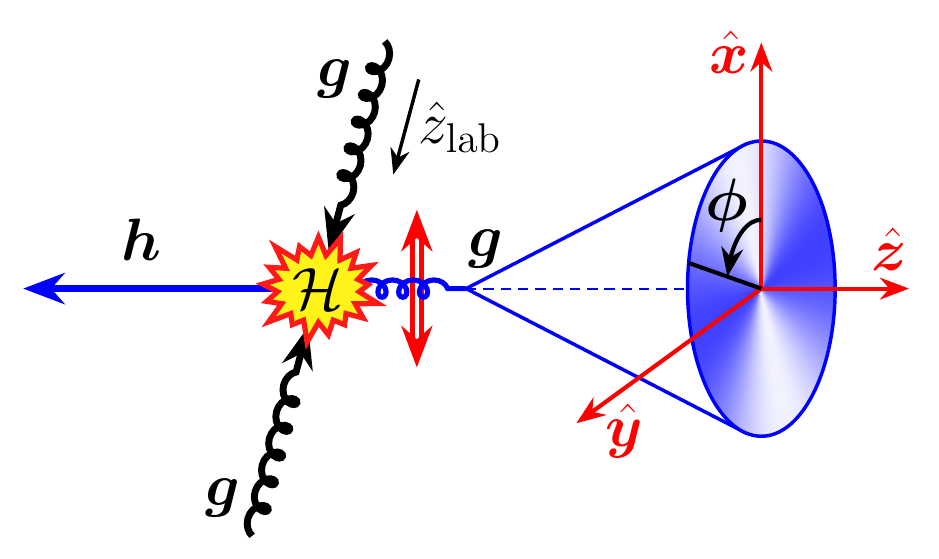}
	\vspace{-3mm}
	\caption{Left: a representative diagram for $gg\to hg$ via a box top loop; 
		the others omitted are $s$-, $t$-, and $u$-channel diagrams involving triangular top loops and tri-gluon vertices.
		Right: the gluon $\hat{x}$-$\hat{y}$-$\hat{z}$ frame defined in \eq{eq:frame}.
	}
	\label{fig:hg}
\end{figure}

In this Letter, we propose a new $CP$-odd observable, for probing the $\htt$ interaction,
which originates from a linearly polarized gluon in the associated production of a Higgs boson and gluon jet ($hg$). 
The essential observation is that a singly polarized gluon can be produced from the hard scattering of unpolarized partons.
After its production, the gluon fragments into a jet with some linear polarization that breaks the rotational invariance around the jet direction and orients the jet constituents according to the hippopedal distribution,
\beq[eq:azimuthal distribution]
	{\rm const.} + \xi_1 \cos2\phi + \xi_2 \sin2\phi \, .
\eeq
Here, as will be defined below, $\xi_1$ and $\xi_2$ parameterize the two degrees of freedom of the linear polarization and depend on both the kinematics of the hard process and the $\htt$ couplings, $\kappa$ and $\tilde{\kappa}$. The azimuthal angle $\phi$ is defined in the $\hat{x}$-$\hat{y}$ plane of the coordinate system,
\beq[eq:frame]
	\hat{z} = \frac{\bm{k} }{ \abs{\bm{k}} } \, , \quad
	\hat{y} = \frac{\hat{z}_{\rm lab} \times \hat{z} }{ \abs{\hat{z}_{\rm lab} \times \hat{z}}} \, ,	\quad
	\hat{x} = \hat{y} \times \hat{z} \, ,
\eeq
shown in \fig{fig:hg} (right),
where $\hat{z}_{\rm lab}$ is the beam direction, and $\bm{k}$ is three-momentum of the gluon jet in the partonic center-of-mass (c.m.) frame. 

The azimuthally anisotropic jet image in \eq{eq:azimuthal distribution} can be measured as a new jet substructure observable and provide sensitivity to the $CP$ phase of the $\htt$ interaction. 
In particular, we will show that $\xi_2$ and the associated $\sin2\phi$ structure are $CP$-odd.
They are more sensitive to a small $CP$ phase $\alpha$ than $\xi_1$, including the sign of $\alpha$.
Contrary to building upon a neutral state of charged particles and antiparticles~\cite{Han:2009ra},
this $CP$-odd observable is constructed purely out of the kinematic information in the gluon jet.
Such $CP$ sensitivity in $hg$ production would not be possible 
without exploring the gluon jet substructure, which has not been considered previously in the literature.
We also note that associated Higgs-top production and indirect measurements via $hV$ or $VV$ production also depend on the $hVV$ couplings and require assumptions on the latter, whereas $hg$ production only depends on the $\htt$ coupling.


\emph{Linearly polarized gluon in $hg$ production.}---The polarization state of the produced gluon is described 
by a density matrix, represented in the helicity basis as
\beq
	\rho_{\lambda\lambda'} 
		= \frac{1}{2}\pp{ 1 + \bm{\xi}\cdot\bm{\sigma} }_{\lambda\lambda'}
		= \frac{1}{2} 
			\begin{pmatrix}
				1 + \xi_3 & \xi_1 - i \, \xi_2	\\
				\xi_1 + i \, \xi_2 & 1 - \xi_3
			\end{pmatrix},
\eeq
with three polarization degrees of freedom, $\bm{\xi} \equiv (\xi_1$, $\xi_2$, $\xi_3)$. 
The diagonal element $\xi_3 = \rho_{++} - \rho_{--}$ describes the net gluon helicity,
whereas the off-diagonal
$\xi_1 = 2\Re\rho_{+-}$ and $\xi_2 = -2\Im\rho_{+-}$ are associated with
interference between the gluon $+$ and $-$ helicity states.
They are better understood in terms of the linear polarization state $|\phi\rangle$
in the $\hat{x}$-$\hat{y}$-$\hat{z}$ frame, related to the helicity eigenstates $|\pm\rangle$ by 
$|\phi\rangle = [ \, e^{i\phi} |-\rangle - e^{-i \phi} |+\rangle \, ] / \sqrt{2}$,
\begin{align}\label{eq:linear pol}
	\xi_1 &= \langle \pi/2 \, | \rho |\, \pi/2 \rangle - \langle 0 \, | \rho | \, 0\rangle = \rho_{yy} - \rho_{xx}	\,,	\nn\\
	\xi_2 &= \langle 3\pi/4 \, | \rho |\, 3\pi/4 \rangle - \langle \pi/4 \, | \rho | \, \pi/4\rangle\ .
\end{align}
Thus, $\xi_1$ and $\xi_2$ are 
differences between 
linear polarization degrees along two orthogonal directions.
It is readily seen that under $CP$ transformation, 
$(\xi_1,\xi_2) \to (\xi_1, -\xi_2)$ 
so they are $CP$-even and $CP$-odd, respectively.
The ambiguity in defining $\hat{z}_{\rm lab}$ in \eq{eq:frame} at a $pp$ collider merely implies the change $(\hat{x}, \hat{y}) \to (-\hat{x}, -\hat{y})$,  
which does not affect linear polarization states, contrary to transverse spins of fermions~\cite{Kane:1991bg}.

The gluon produced in the $hg$ process
has a significant linear polarization.
At leading order (LO), both $gg$ fusion and $q\bar{q}$ annihilation contribute via a top loop, as exemplified in \fig{fig:hg} (left) for the $gg$ channel. 
Even though the $q\bar{q}$ channel can also produce a substantially polarized gluon, its contribution to the total cross section is much smaller and will be neglected.
Parametrizing the helicity amplitudes $g(\lambda_1)\, g(\lambda_2)\to h\, g(\lambda_3)$ in the partonic {\rm c.m.}~frame in terms of the gluon's transverse momentum $p_T$, rapidity $y_g$, and azimuthal angle $\phi_g$, we have 
\begin{align}
	&\M_{\lambda_1\lambda_2\lambda_3}(p_T, y_g, \phi_g) = f^{abc} \,e^{i(\lambda_1 - \lambda_2)\phi_g} \\
	&\hspace{2em} \times
			\bb{ \k \, \A_{\lambda_1 \lambda_2 \lambda_3}(p_T, y_g) + i \, \kt \, \At_{\lambda_1\lambda_2\lambda_3}(p_T, y_g) },	\nn
\end{align}
with $f^{abc}$ the color factor, and $\lambda_i$ the gluon helicities.
The $p_T$ and $y_g$ sufficiently determine the Higgs energy, $E_h^2 = m_H^2 + p_T^2 \cosh^2 y_g$, and the partonic {\rm c.m.}~energy $\sqrt{\hat{s}} = p_T \cosh y_g + E_h$, with $m_H$ being the Higgs mass.
$\A$ and $\At$ are the $CP$-even and $CP$-odd helicity amplitudes, respectively, constrained by their $CP$ properties as
\beq\label{eq:CP amp}
	(\A, \, \At)_{-\lambda_1,-\lambda_2,-\lambda_3}(p_T, y_g) 
		= (-\A, +\At)_{\lambda_1\lambda_2\lambda_3}(p_T, y_g).
\eeq

The gluon density matrix is determined through
\beq[eq:amp rho]
	\frac{1}{4 \, N_{c,g}^2} \M_{\lambda_1 \lambda_2 \lambda} \M^{*}_{\lambda_1 \lambda_2 \lambda'}
		\equiv \rho_{\lambda\lambda'}(\bm{\xi}) \, \abs{\M}^2,
\eeq
where the convention of summing over repeated indices is taken, and $\abs{\M}^2$ is the unpolarized squared amplitude, averaged/summed over the spins and colors, with $N_{c, g} = 8$.
Due to their $CP$ properties in \eq{eq:CP amp}, $\A$ and $\At$ individually only contribute to $\xi_1$, 
while it is their interference that contributes to $\xi_2$.
In terms of the $CP$ phase $\alpha$, $\xi_1$ and $\xi_2$ can be expressed as
\beq[eq:xi12]
	\xi_1 = \frac{ \omega + \beta_1 \cos2\alpha }{1 + \Delta \cos2\alpha },
	\quad
	\xi_2 = \frac{ \beta_2 \sin2\alpha }{1 + \Delta \cos2\alpha }	\,,
\eeq
where we have defined the polarization parameters
\begin{align*}
	\Delta &= \frac{ |\A|^2 - |\At|^2 }{ |\A|^2 + |\At|^2 }	\,,	\quad
	\omega = \frac{2( \A_+ * \A^*_{-} + \At_+ * \At^*_{-} )}{ |\A|^2 + |\At|^2 }	\,,	\\
	\beta_1 &= \frac{2( \A_+ * \A^*_{-} - \At_+ * \At^*_{-} )}{ |\A|^2 + |\At|^2 }	\,,	\quad
	\beta_2 = \frac{ 4 \Re ( \A_+ * \At^*_{-} )}{  |\A|^2 + |\At|^2 }	\,,
\end{align*}
with the notations
\begin{align*}
	A_+ * B_- \equiv A_{\lambda_1\lambda_2+} B_{\lambda_1\lambda_2-},\quad
	\abs{A}^2 \equiv A_{\lambda_1\lambda_2\lambda_3} \, A^*_{\lambda_1\lambda_2\lambda_3}.
\end{align*}
Parametrizing $\xi_{1,2}$ as in \eq{eq:xi12} clearly shows that the polarization only depends on the $CP$ phase $\alpha$, 
but not on the coupling strength $\k_t$, 
which only controls the event rate.
The helicity polarization $\xi_3$ is also nonzero as $\sqrt{\hat{s}} > 2m_t$,
but its value is generally small compared to $\xi_1$ and $\xi_2$, and will not be discussed in this work.

\begin{figure*}[htbp]
	\centering
		\includegraphics[scale=0.5]{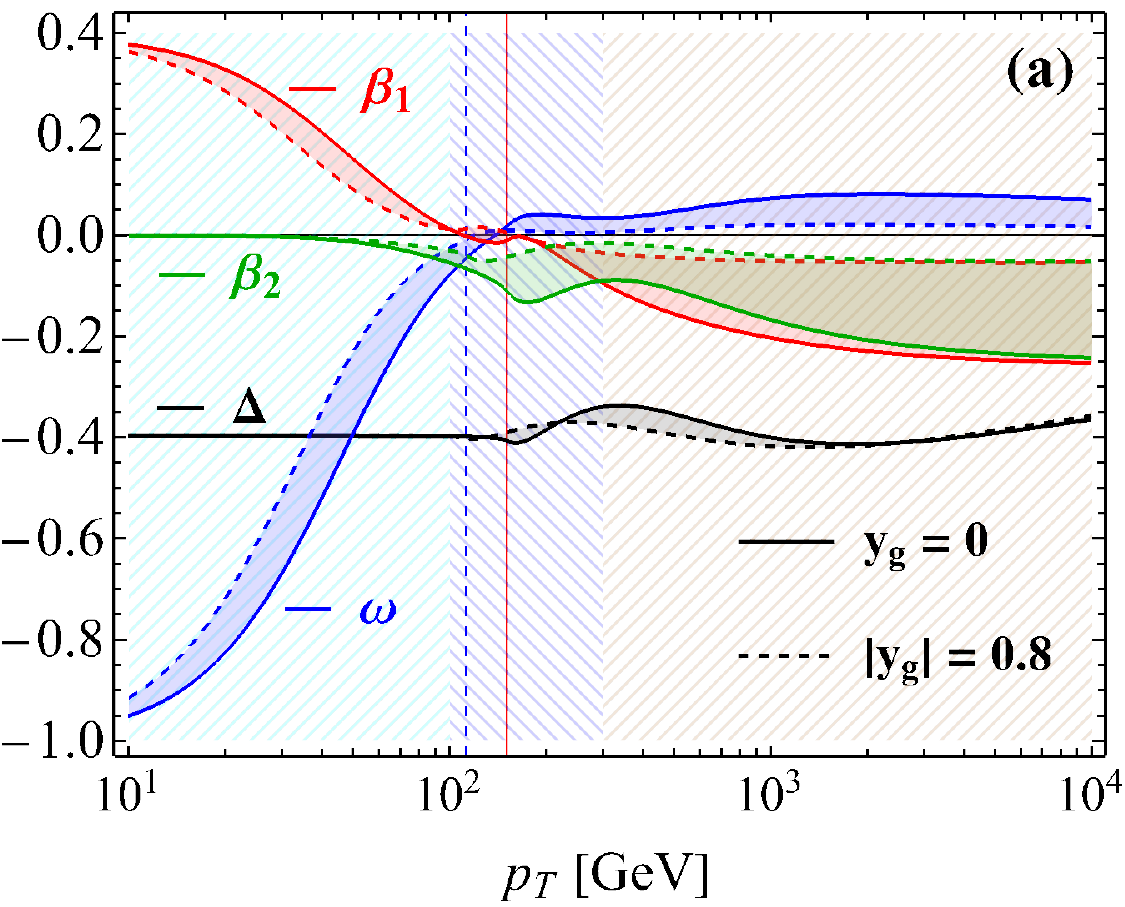} \,
		\includegraphics[scale=0.505]{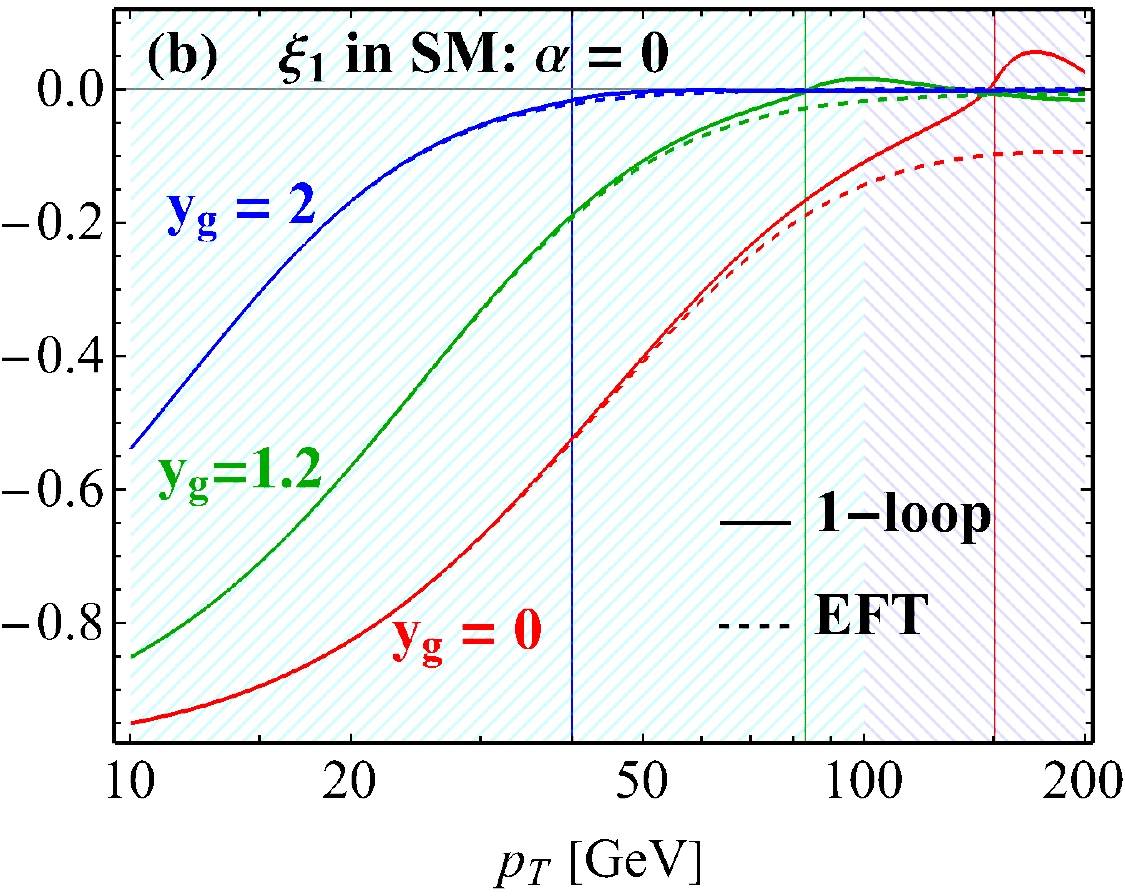} \,
		\includegraphics[scale=0.515]{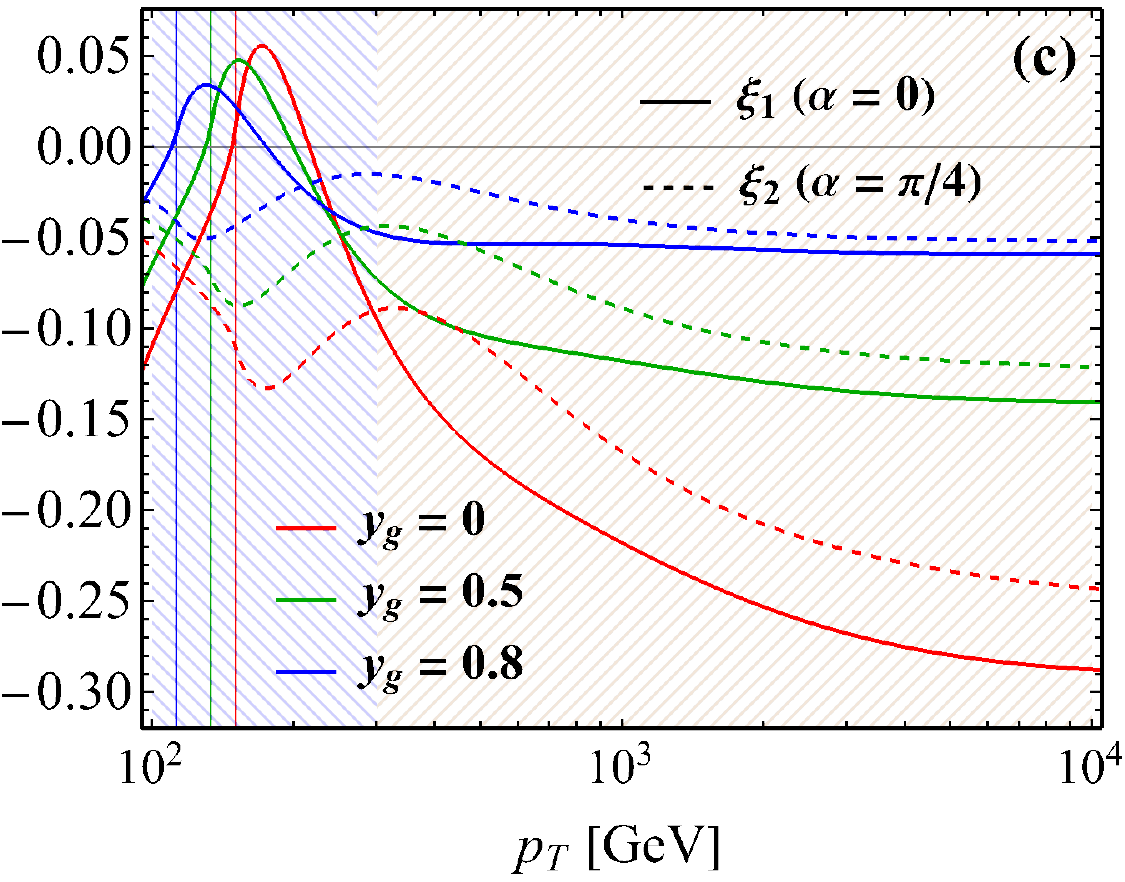} 
	\caption{
	(a) Polarization parameters $\Delta$, $\omega$, $\beta_1$, and $\beta_2$, as functions of the gluon $p_T$ in the partonic {\rm c.m.}~frame. 
	Each parameter is shown as a shaded region constrained by $\abs{y_g} \leq 0.8$, 
	bounded by a solid and a dashed curves, corresponding to $y_g = 0$ and $\abs{y_g} = 0.8$, respectively. 
	The two vertical lines stand for the $\sqrt{\hat{s}} = 2m_t$ threshold for 
	$y_g = 0$ (red, solid) and $\abs{y_g} = 0.8$ (blue, dashed).
	The three hatching-shaded regions are the low-$p_T$ region (cyan) for $p_T < 100~\GeV$, 
	transition region (blue) for $p_T \in (100, 300)~\GeV$, and high-$p_T$ region (brown) for $p_T > 300~\GeV$. 
	(b) $\xi_1$ in the low-$p_T$ region with the SM Lagrangian ($\alpha = 0$) for three values of $y_g$, 
	where the full one-loop calculation (solid) is compared with the EFT result (dashed). 
	The three vertical lines are the $\sqrt{\hat{s}} = 2m_t$ threshold for 
	$y_g = 0$ (red), $y_g = 1.2$ (green) and $y_g = 2$ (blue). 
	(c) $\xi_1$ and $\xi_2$ in the transition and high-$p_T$ regions, 
	for $CP$ phase $\alpha = 0$ and $\pi/4$, respectively.
	}
	\label{fig:pols}
\end{figure*}

The parameters $(\Delta, \, \omega, \, \beta_1, \, \beta_2)$ are functions of $p_T$ and $y_g$, as shown in \fig{fig:pols}(a) 
for some benchmark phase-space points.
While $\Delta$ describing the relative difference between the $CP$-even and $CP$-odd amplitude squares
stays relatively flat around $-0.4$ when $p_T < 10~\TeV$, 
the parameters $\omega$, $\beta_1$, and $\beta_2$, 
which control the sizes of the polarizations $\xi_1$ and $\xi_2$, 
vary sizably with $p_T$.
Based on their $p_T$ dependence, we divide the phase space into three 
kinematic regions and discuss them in turn.

\emph{1. Low-$p_T$ region: $p_T \lesssim 100~\GeV$.}
Both $\abs{\omega}$ and $\beta_1$ have large values, whereas $\beta_2 \simeq 0$. 
The linear polarization is thus dominated by $\xi_1$, with $\xi_2 \simeq 0$. 
The dominance of $\omega$ over $\beta_1$ further implies that $\xi_1$ does not depend sensitively on $\alpha$. 
Being well below the $\sqrt{\hat{s}} = 2m_t$ threshold, this region 
can be well approximated by the infinite-top-mass effective field theory (EFT)~\cite{Dawson:1990zj, Djouadi:1991tka}. 
In \fig{fig:pols}(b), the SM predictions for $\xi_1$ are shown for both the full one-loop calculation and the EFT approximation, where one can see that $\xi_1$ generally has a large negative value, which means that the produced gluon is dominantly polarized along the $\hat{x}$ direction in the production plane, cf.~\eq{eq:linear pol}.
Furthermore, it is not dramatically dependent on the gluon rapidity $y_g$.

Since the low-$p_T$ region contains most of the $hg$ events, it is suitable for testing the linear polarization phenomenon. 
We expect a significant $\cos2\phi$ jet anisotropy due to the dominant $\xi_1$.
The insensitivity to $\alpha$ also enables this region
to serve as a calibration region for experimentally measuring the linear polarization, which is important to ensure its viability and to understand the systematic uncertainties of the measurement since such phenomenon has not been observed before.

\emph{2. Transition region: $100~\GeV \lesssim p_T \lesssim 300~\GeV$.}
$\beta_1$ and $\omega$ rapidly go to 0 and flip their signs, while $\abs{\beta_2}$ starts growing to an appreciable value. 
Hence, the linear polarization is dominated by $\xi_2$ if $\alpha$ is not too small,
as illustrated in \fig{fig:pols}(c) for $\xi_1$ at $\alpha = 0$, and $\xi_2$ at $\alpha = \pi/4$, which corresponds to a maximal $CP$ mixing. 
A nonzero $\alpha$ would then lead to linearly polarized gluon jets featuring a $\sin2\phi$ anisotropy,
whose measurement provides a good opportunity for constraining the $CP$-odd coupling. 
Both $\xi_1$ and $\xi_2$ are sensitive to $y_g$, and their magnitudes are larger for gluons at more central rapidity region.
Since this region covers the $\sqrt{\hat{s}} = 2m_t$ threshold,
EFT is no longer a good approximation, as indicated in the right half of \fig{fig:pols}(b). 

\emph{3. High-$p_T$ region: $p_T \gtrsim 300~\GeV$.}
Both $\beta_1$ and $\beta_2$ have appreciable negative magnitude. 
Their values grow and approach each other as $p_T$ increases. 
Moreover, $\omega$, being smaller than $\abs{\beta_1}$, becomes less important in $\xi_1$. 
Qualitatively, we can interpret this region by taking
$(\omega, \Delta) \to 0$ and $(\beta_1, \beta_2) \to \beta$, 
which gives
$(\xi_1, \xi_2) \sim \beta (\cos2\alpha, \sin2\alpha)$.
Then the jet anisotropy in \eq{eq:azimuthal distribution} can be recast as
\beq[eq:oscillation]
	{\rm const.} + 
	\beta \cos2(\phi - \alpha) \,,
\eeq
so that the main axis direction of the jet image gives a direct measure of the $CP$ phase. 
In fact, it can be shown that as $\hat{s} \to \infty$, this qualitative simplification becomes exact within one-loop calculation.
The quantitative behavior of $\xi_1$ and $\xi_2$ in the high-$p_T$ region is shown in the right half of \fig{fig:pols}(c), where we see that they drop rapidly to 0 as $\abs{y_g}$ increases, and a simple kinematic cut $\abs{y_g} < 0.8$ yields the polarization $\abs{\beta_{1,2}} \gtrsim 0.05$.


\emph{Polarized gluon jet function.}---Around the same time as QCD was developed, 
it was noted that linearly polarized gluons with nonzero $\xi_1$ can be produced in hard collision processes
~\cite{Brodsky:1978hw, Olsen:1979fp, Devoto:1979jm, Devoto:1979fq, DeGrand:1980yp, Petersson:1980cn, 
Koller:1980fk, Olsen:1981ws, Devoto:1981vh, Korner:1981qj, Olsen:1983mm, Hara:1988uj, Robinett:1990qt, 
Jacobsen:1990jp, Groote:1997vg, Groote:1998qt, Groote:2002qc}, 
and some non-perturbative arguments were used in favor of oblate gluon jets characterized by a $\cos2\phi$ distribution. 
In the presence of a $CP$-violating interaction as considered in this work, 
a nonzero $\xi_2$ polarization is also produced leading to an additional $\sin2\phi$ structure, which serves as a handle to probe the $CP$ structure.

Here, we introduce the polarized gluon jet in terms of the modern factorization formalism, for the first time. The polarized gluon turns into a jet that imprints its polarization information in the azimuthal distribution of its constituents, 
which can be projected out by weighting each event by some azimuth-sensitive observable. 
The azimuthally weighted cross section $\sigma_w$ of the inclusive $hg$ production at a $pp$ collider
can be factorized into a hard scattering cross section, 
as given in \eq{eq:amp rho}, multiplied by a polarized gluon jet function, in much the same way as the factorization for an unpolarized jet function~\cite{Berger:2003iw, Almeida:2008tp, Almeida:2008yp} or fragmentation function~\cite{Nayak:2005rt, Collins:2011zzd}. 
It reads as
\begin{align}\label{eq:factorization}
	\frac{ d\sigma_{w} }{dy_g \, dp_T^2 \, d m_J^2 \, d \phi}
			&= \frac{d\hat{\sigma}}{dy_g \, dp_T^2} \, \frac{d J(\bm{\xi}(p_T, y_g), m_J^2, \phi)}{d \phi} \,,
\end{align}
up to corrections of powers of $m_J/p_T$ and the jet size $R$. 
Here, 
$d\hat{\sigma} / dy_g \, dp_T^2 = \lum(s, \hat{s}) \, \abs{\M}^2 / 16\pi E_h \sqrt{\hat{s}}$ 
is the differential cross section for the on-shell gluon production, 
where
$\lum(s, \hat{s}) = \int_{\hat{s}/s}^1 dx/(x s) \, f_{g/{\rm p}}(x, \, \mu_F) f_{g/{\rm p}}(\hat{s}/x s, \, \mu_F)$ 
is the gluon-gluon parton luminosity, 
with the factorization scale chosen at $\mu_F = p_T$ 
in the parton distribution function (PDF) $f_{g/{\rm p}}(x, \, \mu_F)$ of the proton,
and we have used the LO kinematics to integrate over the Higgs phase space.

In the partonic {\rm c.m.}~frame, the gluon momentum $k$ defines the jet mass $m_J^2 = k^2$ and direction $\hat{z}$ as in \eq{eq:frame}.
By defining two lightlike vectors $n^{\mu} = (1, -\hat{z})/\sqrt{2}$ and $\bar{n}^{\mu} = (1, \hat{z})/\sqrt{2}$, we can 
approximate the gluon momentum in the hard part to be on shell by only retaining the large component, $p_g^{\mu} = (k\cdot n) \bar{n}^{\mu}$,
which then defines the rapidity $y_g$ and $p_T = k\cdot n / (\sqrt{2} \cosh y_g)$.
To the leading power of $m_J/p_T$, the on-shell gluon carries the polarization $\bm{\xi}$ and fragments into a jet, 
described by the polarized jet function $dJ(\bm{\xi}, m_J^2, \phi) / d\phi$,
\begin{align}\label{eq:jet}
	\frac{d J}{d \phi}
		&= \frac{1}{2\pi N_{c,g}(k \cdot n)^2} \sum_{X} \int d^4 x \, e^{i k\cdot x}  \bb{ \rho_{\lambda\lambda'}(\bm{\xi})  O(\phi, X) } \nn\\
			&\hspace{2em} \times \varepsilon_{\lambda' \nu}^*(p_g)  \, \langle 0 | W_{ac}(\infty, x; n) \, n_{\sigma} G_c^{\sigma \nu}(x) | X \rangle
				\nn\\
			&\hspace{2em} \times 
			 \varepsilon_{\lambda \mu}(p_g) \, \langle X | W_{ab}(\infty, 0; n) \, n_{\rho} G_b^{\rho \mu}(0) | 0 \rangle 	\,,	
\end{align}
where $X$ denotes the state of the particles within the jet, in accordance with the jet algorithm~\cite{Almeida:2008tp, Ellis:2010rwa}, whose momenta are dominantly along $\bar{n}$. 
$G_c^{\mu\nu}$ is the gluon field strength tensor, and $W_{ab}(\infty, x; n)$ is the Wilson line in the adjoint representation from $x$ to $\infty$ along $n$, 
with the color indices $a, b$, and $c$ summed over.
In \eq{eq:jet}, the gluon polarization states are projected using the on-shell polarization vectors 
$\varepsilon_{\lambda}^{\mu}(p_g)$ with helicity $\lambda = \pm 1$,
which are then averaged with the density matrix $\rho_{\lambda\lambda'}(\bm{\xi})$.
The resultant azimuthal distribution is extracted by inserting the observable
\beq[eq:O]
	O(\phi, X) = \frac{1}{\sum_{i\in X} p_{i,T}} \sum_{i\in X} p_{i,T} \delta(\phi - \phi_i),
\eeq
where $p_{i,T}$ and $\phi_i$ are, respectively, the transverse momentum and azimuthal angle of the jet constituent $i$ with respect to the $\hat{x}$-$\hat{y}$ plane defined in \eq{eq:frame}.
The $\phi$ distribution is a new jet substructure observable introduced by the linear polarization. 
The dependence on $\xi_3$ would vanish due to parity invariance of $O(\phi, X)$.

As a result of the $p_{i,T}$ weight, the observable $O(\phi, X)$ is infrared (IR) safe, and hence
the polarized gluon jet function is insensitive to hadronization effects and becomes perturbatively calculable, with a predictable $\phi$ dependence. 
However,
it was noted long before~\cite{DeGrand:1980yp, Hara:1988uj} that the gluon polarization information will be greatly washed out by the cancellation between the $g\to gg$ and $g\to q\bar{q}$ channels, which was also found recently in a similar situation~\cite{Chen:2020adz, Chen:2021gdk, Larkoski:2022lmv}. It is possible to mitigate these effects by using 
jet flavor tagging techniques~\cite{Gallicchio:2011xq, Gallicchio:2012ez, FerreiradeLima:2016gcz, Frye:2017yrw, Banfi:2006hf, Gras:2017jty, Metodiev:2018ftz, Larkoski:2014pca, Bhattacherjee:2015psa, Kasieczka:2018lwf, Larkoski:2019nwj, Bright-Thonney:2022xkx}.
For example, one may recluster the identified gluon jet into two subjets, and only keep those gluon jets with their two subjets tagged as quarks.
At $\order{\alpha_s}$, requiring a tagged quark in the gluon jet leaves $g \to q\bar{q}$ as the only diagram, 
giving the polarized gluon jet function,
\begin{align}\label{eq:jet q}
	\frac{d J^{(q)}}{d\phi} 
		= \frac{\alpha_s T_F}{6 \pi^2 m_J^2}
			\bb{ 1 + \frac{1}{2} \pp{ \xi_1 \cos2\phi + \xi_2 \sin2\phi  } },
\end{align}
where the jet algorithm dependence does not come in at this order to the leading power of $m_J$.
\eq{eq:jet q} needs to be multiplied by the tagging efficiency when used in \eq{eq:factorization}.
Although flavor tagging reduces the statistics significantly, it enhances the gluon spin analyzing power from $\order{1\%}$ to about $50\%$~\cite{Hara:1988uj} and will improve the statistical precision.

Before closing this section, we note the difference of the gluon polarization from a quark. 
While a transversely polarized light (massless) quark can also be produced from hard scattering processes, 
its transverse spin cannot be conveyed via the {\it perturbative} quark jet function 
due to the chiral symmetry of a massless quark. 
It is hence related to chiral symmetry breaking and must require the presence of 
some non-perturbative functions~\cite{Collins:1993kq, Collins:2011zzd, Kang:2020xyq}.


\emph{Phenomenology.}---The gluon jet azimuthal anisotropy in \eq{eq:jet q} can be experimentally measured by simply constructing the asymmetry observables~\cite{Yu:2021zmw}
\begin{align}
	A_{i} &= \frac{\int_0^{2\pi} d\phi  \, (d\sigma_w / d\phi) \cdot \sgn[ F_i(\phi) ]}{ \int_0^{2\pi} d\phi \, (d\sigma_w / d\phi)} = \frac{\xi_{i}}{\pi},
\label{eq:asym}	
\end{align}
where $i \in \{1,2\}$, $F_1(\phi) = \cos2\phi$ and $F_2(\phi) = \sin 2\phi$.
The uncertainties of the asymmetries $A_{1,2}$ are dominated by statistical ones, given by 
$1/\sqrt{N}$ with 
$N$ being the number of the observed events. 
Now we provide a simple demonstration of the constraining power of the gluon linear polarization on the $CP$ phase, by 
confining ourselves to the transition region 
for both the HL-LHC at $14~\TeV$ and FCC-hh at $100~\TeV$, with integrated luminosities $3~\ab^{-1}$ and $20~\ab^{-1}$, respectively.

The $hg$ cross section in the transition region is estimated 
for the Lagrangian [\eq{eq:L}] 
using CT18NNLO PDFs~\cite{Hou:2019efy}
with {\tt MG5\_aMC@NLO 2.6.7}~\cite{Alwall:2014hca} by first generating the $hg$ events with $p_T \in [100, 300]~\GeV$ and $\abs{\eta_g} \le 2.5$ in the lab frame, and then boosting to the partonic {\rm c.m.}~frame 
with a further cut $\abs{y_g} \le 0.8$, which gives 
$\k_t^2 (0.57 \cos^2\alpha + 1.3 \sin^2\alpha)~\pb$ for the HL-LHC and 
$\k_t^2 (13.7  \cos^2\alpha + 30.7 \sin^2\alpha)~\pb$ for the FCC-hh.
While both $\k_t$ and $\alpha$ affect the total production rate and can be constrained by the measurement of the latter, only $\alpha$ determines the polarization. In the following, we take $\k_t = 1$ and consider the constraint on $\alpha$ from the polarization data.

We are interested in final states where the (fat) gluon jet is composed of a pair of quark subjets. 
While it is possible to also discriminate light quark subjets from gluon subjets,
here we only provide a conservative estimate by 
restricting to the bottom ($b$) and charm ($c$) quark tagging as
used in experiments~\cite{CMS-PAS-BTV-15-002, ATLAS-CONF-2016-002, CMS:2017wtu, ATLAS:2018sgt, ATLAS:2018mgv, ATLAS:2019bwq, ATLAS:2019lwq, CMS:2021scf, ATLAS:2021cxe, ATL-PHYS-PUB-2022-027, ATL-PHYS-PUB-2022-010}.
We estimate the branching fraction $f_{g_{b\bar b}}$ ($f_{g_{c\bar c}}$) of $g\to b\bar{b}$ ($g\to c\bar{c}$)
through parton shower simulation using \texttt{Pythia 8.307}~\cite{Sjostrand:2014zea}, 
which gives $f_{g_{b\bar b}} = 0.013$ and $f_{g_{c\bar c}} = 0.019$ in the selected kinematic region. 
Following Refs.~\cite{ATLAS:2019bwq,ATLAS:2018mgv}, 
we take $b$-tagging efficiency $\epsilon_b = 0.7$ and $c$-tagging efficiency $\epsilon_c = 0.3$.
We consider the diphoton decay channel of the SM Higgs boson and assume a Higgs tagging efficiency $\epsilon_h = 0.002$.
This then gives about  
$(51\cos^2\alpha + 115 \sin^2\alpha)$ reconstructed events at the HL-LHC and 
$(8100\cos^2\alpha + 18200\sin^2\alpha)$ events at the FCC-hh.

\begin{figure}[htbp]
	\centering
		\includegraphics[trim={0 0.1cm -0.5cm 0}, clip, scale=0.5]{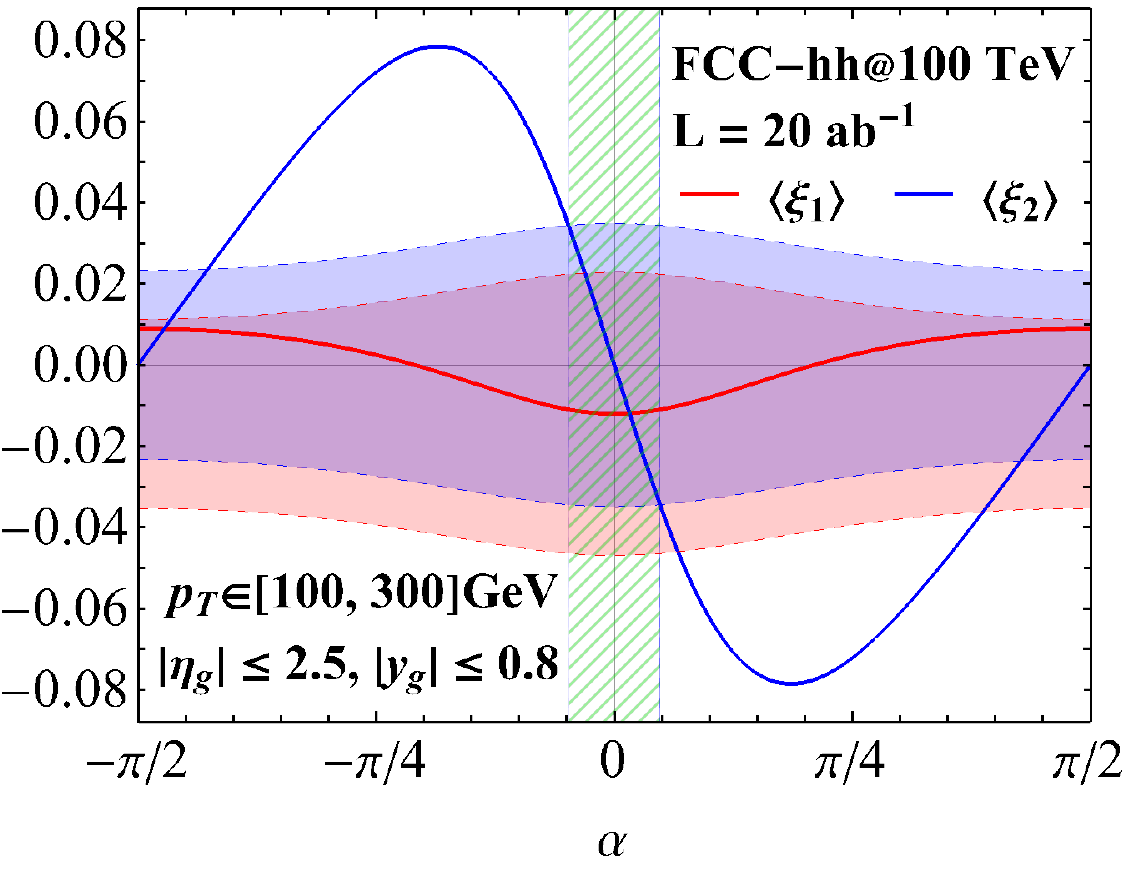} 
	\caption{Constraining power of the FCC-hh gluon polarization data, in the transition region, on the $CP$ phase $\alpha$. $\langle \xi_{1,2} \rangle$ are the average values of $\xi_{1,2}$ in the specified kinematic region.
	Their statistical uncertainties are indicated by the red and blue bands, respectively, around the SM prediction (with $\alpha = 0$).
	The green-hatched region is the $\alpha$ range allowed by the $\xi_2$ measurement.
}
	\label{fig:fcc}
\end{figure}

In \fig{fig:fcc}, we display the predicted average values of $\xi_{1,2}$ in the transition region at the FCC-hh as functions of the $CP$ phase $\alpha$, 
together with their uncertainty bands around the SM central values.
As expected, it is $\xi_2$ that constrains small values of $\alpha$, whereas $\xi_1$ is too small to have an impact in this region.
Assuming the SM scenario with $\xi_2 = 0$, we can project the constraint $\abs{\alpha} \le 8.6^\circ$.
In this estimate, we have only used the gluon polarization information with Higgs decaying to diphotons. 
In order to make a significant impact with data from the HL-LHC, 
one will have to include other Higgs decay channels and 
light quark flavor tagging in the gluon jets, as well as data from the low-$p_T$ and high-$p_T$ regions, 
which shall significantly improve the constraints. 
To the best of our knowledge, however, none of the event generators currently available in public is capable of correctly generating parton showers for a polarized gluon produced directly from the hard part, stopping us from having a more careful phenomenological study, which we must leave for the future. Nevertheless, the anisotropy substructure of the $g\to b \bar{b}$ splitting studied at ATLAS~\cite{ATLAS:2018zhf} should render such observables more hopeful and realistic.


\emph{Summary.}---A precise understanding of the $CP$ property of the Higgs boson is important both to test the SM and to probe new physics. In this Letter, we proposed a novel way of probing the $CP$ structure of the Higgs-top interaction, by measuring the azimuthal anisotropy substructure of the gluon  jet produced in association with a Higgs boson, which originates from the linear polarization of the final-state gluon. We have introduced a factorization formalism and defined a perturbative polarized gluon jet function with 
insertion of an IR-safe azimuthal observable. 
Experimental measurement of the linearly polarized gluon jet will be an important test of the SM and can also serve as a new tool to search for new physics.


\vspace{3mm}
\emph{Acknowledgments.}---We thank B.~Agarwal, J.~Huston, Z.-b.~Kang, A.~J.~Larkoski, A.~v.~Manteuffel, V.~Sotnikov, D.~Stump, J.~Thaler, B.~Yan, T.~Z.~Yang, and H.~X.~Zhu for helpful discussions. This work is in part supported by the U.S.~National Science Foundation under Grant No.~PHY-2013791 and~PHY-2310291. CPY is also grateful for the support from the Wu-Ki Tung endowed chair in particle physics.

\bibliographystyle{apsrev}
\bibliography{reference}

\begin{thebibliography}{94}
\expandafter\ifx\csname natexlab\endcsname\relax\def\natexlab#1{#1}\fi
\expandafter\ifx\csname bibnamefont\endcsname\relax
  \def\bibnamefont#1{#1}\fi
\expandafter\ifx\csname bibfnamefont\endcsname\relax
  \def\bibfnamefont#1{#1}\fi
\expandafter\ifx\csname citenamefont\endcsname\relax
  \def\citenamefont#1{#1}\fi
\expandafter\ifx\csname url\endcsname\relax
  \def\url#1{\texttt{#1}}\fi
\expandafter\ifx\csname urlprefix\endcsname\relax\def\urlprefix{URL }\fi
\providecommand{\bibinfo}[2]{#2}
\providecommand{\eprint}[2][]{\url{#2}}

\bibitem[{\citenamefont{Sirunyan et~al.}(2021)}]{CMS:2021nnc}
\bibinfo{author}{\bibfnamefont{A.~M.} \bibnamefont{Sirunyan}}
  \bibnamefont{et~al.} (\bibinfo{collaboration}{CMS}), \bibinfo{journal}{Phys.
  Rev. D} \textbf{\bibinfo{volume}{104}}, \bibinfo{pages}{052004}
  (\bibinfo{year}{2021}), \eprint{2104.12152}.

\bibitem[{\citenamefont{Sirunyan et~al.}(2020)}]{CMS:2020cga}
\bibinfo{author}{\bibfnamefont{A.~M.} \bibnamefont{Sirunyan}}
  \bibnamefont{et~al.} (\bibinfo{collaboration}{CMS}), \bibinfo{journal}{Phys.
  Rev. Lett.} \textbf{\bibinfo{volume}{125}}, \bibinfo{pages}{061801}
  (\bibinfo{year}{2020}), \eprint{2003.10866}.

\bibitem[{\citenamefont{Aad et~al.}(2020)}]{ATLAS:2020ior}
\bibinfo{author}{\bibfnamefont{G.}~\bibnamefont{Aad}} \bibnamefont{et~al.}
  (\bibinfo{collaboration}{ATLAS}), \bibinfo{journal}{Phys. Rev. Lett.}
  \textbf{\bibinfo{volume}{125}}, \bibinfo{pages}{061802}
  (\bibinfo{year}{2020}), \eprint{2004.04545}.

\bibitem[{\citenamefont{Tumasyan et~al.}(2023)}]{CMS:2022dbt}
\bibinfo{author}{\bibfnamefont{A.}~\bibnamefont{Tumasyan}} \bibnamefont{et~al.}
  (\bibinfo{collaboration}{CMS}), \bibinfo{journal}{JHEP}
  \textbf{\bibinfo{volume}{07}}, \bibinfo{pages}{092} (\bibinfo{year}{2023}),
  \eprint{2208.02686}.

\bibitem[{ATL(2023)}]{ATLAS:2023cbt}
 (\bibinfo{year}{2023}), \eprint{2303.05974}.

\bibitem[{\citenamefont{Sakharov}(1967)}]{Sakharov:1967dj}
\bibinfo{author}{\bibfnamefont{A.~D.} \bibnamefont{Sakharov}},
  \bibinfo{journal}{Pisma Zh. Eksp. Teor. Fiz.} \textbf{\bibinfo{volume}{5}},
  \bibinfo{pages}{32} (\bibinfo{year}{1967}).

\bibitem[{\citenamefont{Ellis et~al.}(2014)\citenamefont{Ellis, Hwang, Sakurai,
  and Takeuchi}}]{Ellis:2013yxa}
\bibinfo{author}{\bibfnamefont{J.}~\bibnamefont{Ellis}},
  \bibinfo{author}{\bibfnamefont{D.~S.} \bibnamefont{Hwang}},
  \bibinfo{author}{\bibfnamefont{K.}~\bibnamefont{Sakurai}}, \bibnamefont{and}
  \bibinfo{author}{\bibfnamefont{M.}~\bibnamefont{Takeuchi}},
  \bibinfo{journal}{JHEP} \textbf{\bibinfo{volume}{04}}, \bibinfo{pages}{004}
  (\bibinfo{year}{2014}), \eprint{1312.5736}.

\bibitem[{\citenamefont{Boudjema et~al.}(2015)\citenamefont{Boudjema, Godbole,
  Guadagnoli, and Mohan}}]{Boudjema:2015nda}
\bibinfo{author}{\bibfnamefont{F.}~\bibnamefont{Boudjema}},
  \bibinfo{author}{\bibfnamefont{R.~M.} \bibnamefont{Godbole}},
  \bibinfo{author}{\bibfnamefont{D.}~\bibnamefont{Guadagnoli}},
  \bibnamefont{and} \bibinfo{author}{\bibfnamefont{K.~A.} \bibnamefont{Mohan}},
  \bibinfo{journal}{Phys. Rev. D} \textbf{\bibinfo{volume}{92}},
  \bibinfo{pages}{015019} (\bibinfo{year}{2015}), \eprint{1501.03157}.

\bibitem[{\citenamefont{Buckley and Goncalves}(2016)}]{Buckley:2015vsa}
\bibinfo{author}{\bibfnamefont{M.~R.} \bibnamefont{Buckley}} \bibnamefont{and}
  \bibinfo{author}{\bibfnamefont{D.}~\bibnamefont{Goncalves}},
  \bibinfo{journal}{Phys. Rev. Lett.} \textbf{\bibinfo{volume}{116}},
  \bibinfo{pages}{091801} (\bibinfo{year}{2016}), \eprint{1507.07926}.

\bibitem[{\citenamefont{Gritsan et~al.}(2016)\citenamefont{Gritsan, R\"ontsch,
  Schulze, and Xiao}}]{Gritsan:2016hjl}
\bibinfo{author}{\bibfnamefont{A.~V.} \bibnamefont{Gritsan}},
  \bibinfo{author}{\bibfnamefont{R.}~\bibnamefont{R\"ontsch}},
  \bibinfo{author}{\bibfnamefont{M.}~\bibnamefont{Schulze}}, \bibnamefont{and}
  \bibinfo{author}{\bibfnamefont{M.}~\bibnamefont{Xiao}},
  \bibinfo{journal}{Phys. Rev. D} \textbf{\bibinfo{volume}{94}},
  \bibinfo{pages}{055023} (\bibinfo{year}{2016}), \eprint{1606.03107}.

\bibitem[{\citenamefont{Mileo et~al.}(2016)\citenamefont{Mileo, Kiers,
  Szynkman, Crane, and Gegner}}]{Mileo:2016mxg}
\bibinfo{author}{\bibfnamefont{N.}~\bibnamefont{Mileo}},
  \bibinfo{author}{\bibfnamefont{K.}~\bibnamefont{Kiers}},
  \bibinfo{author}{\bibfnamefont{A.}~\bibnamefont{Szynkman}},
  \bibinfo{author}{\bibfnamefont{D.}~\bibnamefont{Crane}}, \bibnamefont{and}
  \bibinfo{author}{\bibfnamefont{E.}~\bibnamefont{Gegner}},
  \bibinfo{journal}{JHEP} \textbf{\bibinfo{volume}{07}}, \bibinfo{pages}{056}
  (\bibinfo{year}{2016}), \eprint{1603.03632}.

\bibitem[{\citenamefont{Amor Dos~Santos et~al.}(2017)}]{AmorDosSantos:2017ayi}
\bibinfo{author}{\bibfnamefont{S.}~\bibnamefont{Amor Dos~Santos}}
  \bibnamefont{et~al.}, \bibinfo{journal}{Phys. Rev. D}
  \textbf{\bibinfo{volume}{96}}, \bibinfo{pages}{013004}
  (\bibinfo{year}{2017}), \eprint{1704.03565}.

\bibitem[{\citenamefont{Azevedo et~al.}(2018)\citenamefont{Azevedo, Onofre,
  Filthaut, and Gon\c{c}alo}}]{Azevedo:2017qiz}
\bibinfo{author}{\bibfnamefont{D.}~\bibnamefont{Azevedo}},
  \bibinfo{author}{\bibfnamefont{A.}~\bibnamefont{Onofre}},
  \bibinfo{author}{\bibfnamefont{F.}~\bibnamefont{Filthaut}}, \bibnamefont{and}
  \bibinfo{author}{\bibfnamefont{R.}~\bibnamefont{Gon\c{c}alo}},
  \bibinfo{journal}{Phys. Rev. D} \textbf{\bibinfo{volume}{98}},
  \bibinfo{pages}{033004} (\bibinfo{year}{2018}), \eprint{1711.05292}.

\bibitem[{\citenamefont{Li et~al.}(2018)\citenamefont{Li, Si, Wu, and
  Yue}}]{Li:2017dyz}
\bibinfo{author}{\bibfnamefont{J.}~\bibnamefont{Li}},
  \bibinfo{author}{\bibfnamefont{Z.-g.} \bibnamefont{Si}},
  \bibinfo{author}{\bibfnamefont{L.}~\bibnamefont{Wu}}, \bibnamefont{and}
  \bibinfo{author}{\bibfnamefont{J.}~\bibnamefont{Yue}},
  \bibinfo{journal}{Phys. Lett. B} \textbf{\bibinfo{volume}{779}},
  \bibinfo{pages}{72} (\bibinfo{year}{2018}), \eprint{1701.00224}.

\bibitem[{\citenamefont{Gon\c{c}alves et~al.}(2018)\citenamefont{Gon\c{c}alves,
  Kong, and Kim}}]{Goncalves:2018agy}
\bibinfo{author}{\bibfnamefont{D.}~\bibnamefont{Gon\c{c}alves}},
  \bibinfo{author}{\bibfnamefont{K.}~\bibnamefont{Kong}}, \bibnamefont{and}
  \bibinfo{author}{\bibfnamefont{J.~H.} \bibnamefont{Kim}},
  \bibinfo{journal}{JHEP} \textbf{\bibinfo{volume}{06}}, \bibinfo{pages}{079}
  (\bibinfo{year}{2018}), \eprint{1804.05874}.

\bibitem[{\citenamefont{Faroughy et~al.}(2020)\citenamefont{Faroughy, Kamenik,
  Ko\v{s}nik, and Smolkovi\v{c}}}]{Faroughy:2019ird}
\bibinfo{author}{\bibfnamefont{D.~A.} \bibnamefont{Faroughy}},
  \bibinfo{author}{\bibfnamefont{J.~F.} \bibnamefont{Kamenik}},
  \bibinfo{author}{\bibfnamefont{N.}~\bibnamefont{Ko\v{s}nik}},
  \bibnamefont{and}
  \bibinfo{author}{\bibfnamefont{A.}~\bibnamefont{Smolkovi\v{c}}},
  \bibinfo{journal}{JHEP} \textbf{\bibinfo{volume}{02}}, \bibinfo{pages}{085}
  (\bibinfo{year}{2020}), \eprint{1909.00007}.

\bibitem[{\citenamefont{Bortolato et~al.}(2021)\citenamefont{Bortolato,
  Kamenik, Ko\v{s}nik, and Smolkovi\v{c}}}]{Bortolato:2020zcg}
\bibinfo{author}{\bibfnamefont{B.}~\bibnamefont{Bortolato}},
  \bibinfo{author}{\bibfnamefont{J.~F.} \bibnamefont{Kamenik}},
  \bibinfo{author}{\bibfnamefont{N.}~\bibnamefont{Ko\v{s}nik}},
  \bibnamefont{and}
  \bibinfo{author}{\bibfnamefont{A.}~\bibnamefont{Smolkovi\v{c}}},
  \bibinfo{journal}{Nucl. Phys. B} \textbf{\bibinfo{volume}{964}},
  \bibinfo{pages}{115328} (\bibinfo{year}{2021}), \eprint{2006.13110}.

\bibitem[{\citenamefont{Cao et~al.}(2021)\citenamefont{Cao, Xie, Zhang, and
  Zhang}}]{Cao:2020hhb}
\bibinfo{author}{\bibfnamefont{Q.-H.} \bibnamefont{Cao}},
  \bibinfo{author}{\bibfnamefont{K.-P.} \bibnamefont{Xie}},
  \bibinfo{author}{\bibfnamefont{H.}~\bibnamefont{Zhang}}, \bibnamefont{and}
  \bibinfo{author}{\bibfnamefont{R.}~\bibnamefont{Zhang}},
  \bibinfo{journal}{Chin. Phys. C} \textbf{\bibinfo{volume}{45}},
  \bibinfo{pages}{023117} (\bibinfo{year}{2021}), \eprint{2008.13442}.

\bibitem[{\citenamefont{Gon\c{c}alves et~al.}(2022)\citenamefont{Gon\c{c}alves,
  Kim, Kong, and Wu}}]{Goncalves:2021dcu}
\bibinfo{author}{\bibfnamefont{D.}~\bibnamefont{Gon\c{c}alves}},
  \bibinfo{author}{\bibfnamefont{J.~H.} \bibnamefont{Kim}},
  \bibinfo{author}{\bibfnamefont{K.}~\bibnamefont{Kong}}, \bibnamefont{and}
  \bibinfo{author}{\bibfnamefont{Y.}~\bibnamefont{Wu}}, \bibinfo{journal}{JHEP}
  \textbf{\bibinfo{volume}{01}}, \bibinfo{pages}{158} (\bibinfo{year}{2022}),
  \eprint{2108.01083}.

\bibitem[{\citenamefont{Patrick et~al.}(2020)\citenamefont{Patrick, Scaffidi,
  and Sharma}}]{Patrick:2019nhv}
\bibinfo{author}{\bibfnamefont{R.}~\bibnamefont{Patrick}},
  \bibinfo{author}{\bibfnamefont{A.}~\bibnamefont{Scaffidi}}, \bibnamefont{and}
  \bibinfo{author}{\bibfnamefont{P.}~\bibnamefont{Sharma}},
  \bibinfo{journal}{Phys. Rev. D} \textbf{\bibinfo{volume}{101}},
  \bibinfo{pages}{093005} (\bibinfo{year}{2020}), \eprint{1909.12772}.

\bibitem[{\citenamefont{Hankele et~al.}(2006)\citenamefont{Hankele, Klamke, and
  Zeppenfeld}}]{Hankele:2006ja}
\bibinfo{author}{\bibfnamefont{V.}~\bibnamefont{Hankele}},
  \bibinfo{author}{\bibfnamefont{G.}~\bibnamefont{Klamke}}, \bibnamefont{and}
  \bibinfo{author}{\bibfnamefont{D.}~\bibnamefont{Zeppenfeld}}, in
  \emph{\bibinfo{booktitle}{{Meeting on CP Violation and Non-standard Higgs
  Physics}}} (\bibinfo{year}{2006}), \eprint{hep-ph/0605117}.

\bibitem[{\citenamefont{Brod et~al.}(2013)\citenamefont{Brod, Haisch, and
  Zupan}}]{Brod:2013cka}
\bibinfo{author}{\bibfnamefont{J.}~\bibnamefont{Brod}},
  \bibinfo{author}{\bibfnamefont{U.}~\bibnamefont{Haisch}}, \bibnamefont{and}
  \bibinfo{author}{\bibfnamefont{J.}~\bibnamefont{Zupan}},
  \bibinfo{journal}{JHEP} \textbf{\bibinfo{volume}{11}}, \bibinfo{pages}{180}
  (\bibinfo{year}{2013}), \eprint{1310.1385}.

\bibitem[{\citenamefont{Dolan et~al.}(2014)\citenamefont{Dolan, Harris,
  Jankowiak, and Spannowsky}}]{Dolan:2014upa}
\bibinfo{author}{\bibfnamefont{M.~J.} \bibnamefont{Dolan}},
  \bibinfo{author}{\bibfnamefont{P.}~\bibnamefont{Harris}},
  \bibinfo{author}{\bibfnamefont{M.}~\bibnamefont{Jankowiak}},
  \bibnamefont{and}
  \bibinfo{author}{\bibfnamefont{M.}~\bibnamefont{Spannowsky}},
  \bibinfo{journal}{Phys. Rev. D} \textbf{\bibinfo{volume}{90}},
  \bibinfo{pages}{073008} (\bibinfo{year}{2014}), \eprint{1406.3322}.

\bibitem[{\citenamefont{Englert et~al.}(2013)\citenamefont{Englert,
  Goncalves-Netto, Mawatari, and Plehn}}]{Englert:2012xt}
\bibinfo{author}{\bibfnamefont{C.}~\bibnamefont{Englert}},
  \bibinfo{author}{\bibfnamefont{D.}~\bibnamefont{Goncalves-Netto}},
  \bibinfo{author}{\bibfnamefont{K.}~\bibnamefont{Mawatari}}, \bibnamefont{and}
  \bibinfo{author}{\bibfnamefont{T.}~\bibnamefont{Plehn}},
  \bibinfo{journal}{JHEP} \textbf{\bibinfo{volume}{01}}, \bibinfo{pages}{148}
  (\bibinfo{year}{2013}), \eprint{1212.0843}.

\bibitem[{\citenamefont{Bernlochner et~al.}(2019)\citenamefont{Bernlochner,
  Englert, Hays, Lohwasser, Mildner, Pilkington, Price, and
  Spannowsky}}]{Bernlochner:2018opw}
\bibinfo{author}{\bibfnamefont{F.~U.} \bibnamefont{Bernlochner}},
  \bibinfo{author}{\bibfnamefont{C.}~\bibnamefont{Englert}},
  \bibinfo{author}{\bibfnamefont{C.}~\bibnamefont{Hays}},
  \bibinfo{author}{\bibfnamefont{K.}~\bibnamefont{Lohwasser}},
  \bibinfo{author}{\bibfnamefont{H.}~\bibnamefont{Mildner}},
  \bibinfo{author}{\bibfnamefont{A.}~\bibnamefont{Pilkington}},
  \bibinfo{author}{\bibfnamefont{D.~D.} \bibnamefont{Price}}, \bibnamefont{and}
  \bibinfo{author}{\bibfnamefont{M.}~\bibnamefont{Spannowsky}},
  \bibinfo{journal}{Phys. Lett. B} \textbf{\bibinfo{volume}{790}},
  \bibinfo{pages}{372} (\bibinfo{year}{2019}), \eprint{1808.06577}.

\bibitem[{\citenamefont{Englert et~al.}(2019)\citenamefont{Englert, Galler,
  Pilkington, and Spannowsky}}]{Englert:2019xhk}
\bibinfo{author}{\bibfnamefont{C.}~\bibnamefont{Englert}},
  \bibinfo{author}{\bibfnamefont{P.}~\bibnamefont{Galler}},
  \bibinfo{author}{\bibfnamefont{A.}~\bibnamefont{Pilkington}},
  \bibnamefont{and}
  \bibinfo{author}{\bibfnamefont{M.}~\bibnamefont{Spannowsky}},
  \bibinfo{journal}{Phys. Rev. D} \textbf{\bibinfo{volume}{99}},
  \bibinfo{pages}{095007} (\bibinfo{year}{2019}), \eprint{1901.05982}.

\bibitem[{\citenamefont{Gritsan et~al.}(2020)\citenamefont{Gritsan, Roskes,
  Sarica, Schulze, Xiao, and Zhou}}]{Gritsan:2020pib}
\bibinfo{author}{\bibfnamefont{A.~V.} \bibnamefont{Gritsan}},
  \bibinfo{author}{\bibfnamefont{J.}~\bibnamefont{Roskes}},
  \bibinfo{author}{\bibfnamefont{U.}~\bibnamefont{Sarica}},
  \bibinfo{author}{\bibfnamefont{M.}~\bibnamefont{Schulze}},
  \bibinfo{author}{\bibfnamefont{M.}~\bibnamefont{Xiao}}, \bibnamefont{and}
  \bibinfo{author}{\bibfnamefont{Y.}~\bibnamefont{Zhou}},
  \bibinfo{journal}{Phys. Rev. D} \textbf{\bibinfo{volume}{102}},
  \bibinfo{pages}{056022} (\bibinfo{year}{2020}), \eprint{2002.09888}.

\bibitem[{\citenamefont{Bahl et~al.}(2020)\citenamefont{Bahl, Bechtle,
  Heinemeyer, Katzy, Klingl, Peters, Saimpert, Stefaniak, and
  Weiglein}}]{Bahl:2020wee}
\bibinfo{author}{\bibfnamefont{H.}~\bibnamefont{Bahl}},
  \bibinfo{author}{\bibfnamefont{P.}~\bibnamefont{Bechtle}},
  \bibinfo{author}{\bibfnamefont{S.}~\bibnamefont{Heinemeyer}},
  \bibinfo{author}{\bibfnamefont{J.}~\bibnamefont{Katzy}},
  \bibinfo{author}{\bibfnamefont{T.}~\bibnamefont{Klingl}},
  \bibinfo{author}{\bibfnamefont{K.}~\bibnamefont{Peters}},
  \bibinfo{author}{\bibfnamefont{M.}~\bibnamefont{Saimpert}},
  \bibinfo{author}{\bibfnamefont{T.}~\bibnamefont{Stefaniak}},
  \bibnamefont{and} \bibinfo{author}{\bibfnamefont{G.}~\bibnamefont{Weiglein}},
  \bibinfo{journal}{JHEP} \textbf{\bibinfo{volume}{11}}, \bibinfo{pages}{127}
  (\bibinfo{year}{2020}), \eprint{2007.08542}.

\bibitem[{\citenamefont{Martini et~al.}(2021)\citenamefont{Martini, Pan,
  Schulze, and Xiao}}]{Martini:2021uey}
\bibinfo{author}{\bibfnamefont{T.}~\bibnamefont{Martini}},
  \bibinfo{author}{\bibfnamefont{R.-Q.} \bibnamefont{Pan}},
  \bibinfo{author}{\bibfnamefont{M.}~\bibnamefont{Schulze}}, \bibnamefont{and}
  \bibinfo{author}{\bibfnamefont{M.}~\bibnamefont{Xiao}},
  \bibinfo{journal}{Phys. Rev. D} \textbf{\bibinfo{volume}{104}},
  \bibinfo{pages}{055045} (\bibinfo{year}{2021}), \eprint{2104.04277}.

\bibitem[{\citenamefont{Ren et~al.}(2020)\citenamefont{Ren, Wu, and
  Yang}}]{Ren:2019xhp}
\bibinfo{author}{\bibfnamefont{J.}~\bibnamefont{Ren}},
  \bibinfo{author}{\bibfnamefont{L.}~\bibnamefont{Wu}}, \bibnamefont{and}
  \bibinfo{author}{\bibfnamefont{J.~M.} \bibnamefont{Yang}},
  \bibinfo{journal}{Phys. Lett. B} \textbf{\bibinfo{volume}{802}},
  \bibinfo{pages}{135198} (\bibinfo{year}{2020}), \eprint{1901.05627}.

\bibitem[{\citenamefont{Bahl and Brass}(2022)}]{Bahl:2021dnc}
\bibinfo{author}{\bibfnamefont{H.}~\bibnamefont{Bahl}} \bibnamefont{and}
  \bibinfo{author}{\bibfnamefont{S.}~\bibnamefont{Brass}},
  \bibinfo{journal}{JHEP} \textbf{\bibinfo{volume}{03}}, \bibinfo{pages}{017}
  (\bibinfo{year}{2022}), \eprint{2110.10177}.

\bibitem[{\citenamefont{Barman et~al.}(2022)\citenamefont{Barman,
  Gon\c{c}alves, and Kling}}]{Barman:2021yfh}
\bibinfo{author}{\bibfnamefont{R.~K.} \bibnamefont{Barman}},
  \bibinfo{author}{\bibfnamefont{D.}~\bibnamefont{Gon\c{c}alves}},
  \bibnamefont{and} \bibinfo{author}{\bibfnamefont{F.}~\bibnamefont{Kling}},
  \bibinfo{journal}{Phys. Rev. D} \textbf{\bibinfo{volume}{105}},
  \bibinfo{pages}{035023} (\bibinfo{year}{2022}), \eprint{2110.07635}.

\bibitem[{\citenamefont{Apollinari et~al.}(2017)\citenamefont{Apollinari,
  Brüning, Nakamoto, and Rossi}}]{Apollinari:2120673}
\bibinfo{author}{\bibfnamefont{G.}~\bibnamefont{Apollinari}},
  \bibinfo{author}{\bibfnamefont{O.}~\bibnamefont{Brüning}},
  \bibinfo{author}{\bibfnamefont{T.}~\bibnamefont{Nakamoto}}, \bibnamefont{and}
  \bibinfo{author}{\bibfnamefont{L.}~\bibnamefont{Rossi}},
  \bibinfo{journal}{CERN Yellow Report} pp. \bibinfo{pages}{1--19. 21 p}
  (\bibinfo{year}{2017}), \bibinfo{note}{21 pages, chapter in High-Luminosity
  Large Hadron Collider (HL-LHC) : Preliminary Design Report},
  \eprint{1705.08830}, \urlprefix\url{https://cds.cern.ch/record/2120673}.

\bibitem[{\citenamefont{Mangano and Mangano}(2017)}]{Mangano:2270978}
\bibinfo{author}{\bibfnamefont{M.}~\bibnamefont{Mangano}} \bibnamefont{and}
  \bibinfo{author}{\bibfnamefont{M.}~\bibnamefont{Mangano}},
  \emph{\bibinfo{title}{{Physics at the FCC-hh, a 100 TeV $pp$ collider}}},
  CERN Yellow Reports: Monographs (\bibinfo{publisher}{CERN},
  \bibinfo{address}{Geneva}, \bibinfo{year}{2017}),
  \urlprefix\url{http://cds.cern.ch/record/2270978}.

\bibitem[{\citenamefont{Han and Li}(2010)}]{Han:2009ra}
\bibinfo{author}{\bibfnamefont{T.}~\bibnamefont{Han}} \bibnamefont{and}
  \bibinfo{author}{\bibfnamefont{Y.}~\bibnamefont{Li}}, \bibinfo{journal}{Phys.
  Lett. B} \textbf{\bibinfo{volume}{683}}, \bibinfo{pages}{278}
  (\bibinfo{year}{2010}), \eprint{0911.2933}.

\bibitem[{\citenamefont{Kane et~al.}(1992)\citenamefont{Kane, Ladinsky, and
  Yuan}}]{Kane:1991bg}
\bibinfo{author}{\bibfnamefont{G.~L.} \bibnamefont{Kane}},
  \bibinfo{author}{\bibfnamefont{G.~A.} \bibnamefont{Ladinsky}},
  \bibnamefont{and} \bibinfo{author}{\bibfnamefont{C.~P.} \bibnamefont{Yuan}},
  \bibinfo{journal}{Phys. Rev. D} \textbf{\bibinfo{volume}{45}},
  \bibinfo{pages}{124} (\bibinfo{year}{1992}).

\bibitem[{\citenamefont{Dawson}(1991)}]{Dawson:1990zj}
\bibinfo{author}{\bibfnamefont{S.}~\bibnamefont{Dawson}},
  \bibinfo{journal}{Nucl. Phys. B} \textbf{\bibinfo{volume}{359}},
  \bibinfo{pages}{283} (\bibinfo{year}{1991}).

\bibitem[{\citenamefont{Djouadi et~al.}(1991)\citenamefont{Djouadi, Spira, and
  Zerwas}}]{Djouadi:1991tka}
\bibinfo{author}{\bibfnamefont{A.}~\bibnamefont{Djouadi}},
  \bibinfo{author}{\bibfnamefont{M.}~\bibnamefont{Spira}}, \bibnamefont{and}
  \bibinfo{author}{\bibfnamefont{P.~M.} \bibnamefont{Zerwas}},
  \bibinfo{journal}{Phys. Lett. B} \textbf{\bibinfo{volume}{264}},
  \bibinfo{pages}{440} (\bibinfo{year}{1991}).

\bibitem[{\citenamefont{Brodsky et~al.}(1978)\citenamefont{Brodsky, DeGrand,
  and Schwitters}}]{Brodsky:1978hw}
\bibinfo{author}{\bibfnamefont{S.~J.} \bibnamefont{Brodsky}},
  \bibinfo{author}{\bibfnamefont{T.~A.} \bibnamefont{DeGrand}},
  \bibnamefont{and}
  \bibinfo{author}{\bibfnamefont{R.}~\bibnamefont{Schwitters}},
  \bibinfo{journal}{Phys. Lett. B} \textbf{\bibinfo{volume}{79}},
  \bibinfo{pages}{255} (\bibinfo{year}{1978}).

\bibitem[{\citenamefont{Olsen et~al.}(1980)\citenamefont{Olsen, Osland, and
  Overbo}}]{Olsen:1979fp}
\bibinfo{author}{\bibfnamefont{H.~A.} \bibnamefont{Olsen}},
  \bibinfo{author}{\bibfnamefont{P.}~\bibnamefont{Osland}}, \bibnamefont{and}
  \bibinfo{author}{\bibfnamefont{I.}~\bibnamefont{Overbo}},
  \bibinfo{journal}{Phys. Lett. B} \textbf{\bibinfo{volume}{89}},
  \bibinfo{pages}{221} (\bibinfo{year}{1980}).

\bibitem[{\citenamefont{Devoto et~al.}(1980)\citenamefont{Devoto, Pumplin,
  Repko, and Kane}}]{Devoto:1979jm}
\bibinfo{author}{\bibfnamefont{A.}~\bibnamefont{Devoto}},
  \bibinfo{author}{\bibfnamefont{J.}~\bibnamefont{Pumplin}},
  \bibinfo{author}{\bibfnamefont{W.~W.} \bibnamefont{Repko}}, \bibnamefont{and}
  \bibinfo{author}{\bibfnamefont{G.~L.} \bibnamefont{Kane}},
  \bibinfo{journal}{Phys. Lett. B} \textbf{\bibinfo{volume}{90}},
  \bibinfo{pages}{436} (\bibinfo{year}{1980}).

\bibitem[{\citenamefont{Devoto et~al.}(1979)\citenamefont{Devoto, Pumplin,
  Repko, and Kane}}]{Devoto:1979fq}
\bibinfo{author}{\bibfnamefont{A.}~\bibnamefont{Devoto}},
  \bibinfo{author}{\bibfnamefont{J.}~\bibnamefont{Pumplin}},
  \bibinfo{author}{\bibfnamefont{W.}~\bibnamefont{Repko}}, \bibnamefont{and}
  \bibinfo{author}{\bibfnamefont{G.~L.} \bibnamefont{Kane}},
  \bibinfo{journal}{Phys. Rev. Lett.} \textbf{\bibinfo{volume}{43}},
  \bibinfo{pages}{1062} (\bibinfo{year}{1979}), \bibinfo{note}{[Erratum:
  Phys.Rev.Lett. 43, 1540 (1979)]}.

\bibitem[{\citenamefont{DeGrand and Petersson}(1980)}]{DeGrand:1980yp}
\bibinfo{author}{\bibfnamefont{T.~A.} \bibnamefont{DeGrand}} \bibnamefont{and}
  \bibinfo{author}{\bibfnamefont{B.}~\bibnamefont{Petersson}},
  \bibinfo{journal}{Phys. Rev. D} \textbf{\bibinfo{volume}{21}},
  \bibinfo{pages}{3129} (\bibinfo{year}{1980}).

\bibitem[{\citenamefont{Petersson and Pire}(1980)}]{Petersson:1980cn}
\bibinfo{author}{\bibfnamefont{B.}~\bibnamefont{Petersson}} \bibnamefont{and}
  \bibinfo{author}{\bibfnamefont{B.}~\bibnamefont{Pire}},
  \bibinfo{journal}{Phys. Lett. B} \textbf{\bibinfo{volume}{95}},
  \bibinfo{pages}{119} (\bibinfo{year}{1980}).

\bibitem[{\citenamefont{Koller et~al.}(1981)\citenamefont{Koller, Streng,
  Walsh, and Zerwas}}]{Koller:1980fk}
\bibinfo{author}{\bibfnamefont{K.}~\bibnamefont{Koller}},
  \bibinfo{author}{\bibfnamefont{K.~H.} \bibnamefont{Streng}},
  \bibinfo{author}{\bibfnamefont{T.~F.} \bibnamefont{Walsh}}, \bibnamefont{and}
  \bibinfo{author}{\bibfnamefont{P.~M.} \bibnamefont{Zerwas}},
  \bibinfo{journal}{Nucl. Phys. B} \textbf{\bibinfo{volume}{193}},
  \bibinfo{pages}{61} (\bibinfo{year}{1981}).

\bibitem[{\citenamefont{Olsen et~al.}(1981)\citenamefont{Olsen, Osland, and
  Overbo}}]{Olsen:1981ws}
\bibinfo{author}{\bibfnamefont{H.~A.} \bibnamefont{Olsen}},
  \bibinfo{author}{\bibfnamefont{P.}~\bibnamefont{Osland}}, \bibnamefont{and}
  \bibinfo{author}{\bibfnamefont{I.}~\bibnamefont{Overbo}},
  \bibinfo{journal}{Nucl. Phys. B} \textbf{\bibinfo{volume}{192}},
  \bibinfo{pages}{33} (\bibinfo{year}{1981}).

\bibitem[{\citenamefont{Devoto and Repko}(1982)}]{Devoto:1981vh}
\bibinfo{author}{\bibfnamefont{A.}~\bibnamefont{Devoto}} \bibnamefont{and}
  \bibinfo{author}{\bibfnamefont{W.~W.} \bibnamefont{Repko}},
  \bibinfo{journal}{Phys. Rev. D} \textbf{\bibinfo{volume}{25}},
  \bibinfo{pages}{904} (\bibinfo{year}{1982}).

\bibitem[{\citenamefont{Korner and Schiller}(1981)}]{Korner:1981qj}
\bibinfo{author}{\bibfnamefont{J.~G.} \bibnamefont{Korner}} \bibnamefont{and}
  \bibinfo{author}{\bibfnamefont{D.~H.} \bibnamefont{Schiller}}
  (\bibinfo{year}{1981}).

\bibitem[{\citenamefont{Olsen and Olsen}(1984)}]{Olsen:1983mm}
\bibinfo{author}{\bibfnamefont{O.~E.} \bibnamefont{Olsen}} \bibnamefont{and}
  \bibinfo{author}{\bibfnamefont{H.~A.} \bibnamefont{Olsen}},
  \bibinfo{journal}{Phys. Scripta} \textbf{\bibinfo{volume}{29}},
  \bibinfo{pages}{12} (\bibinfo{year}{1984}).

\bibitem[{\citenamefont{Hara and Sakai}(1989)}]{Hara:1988uj}
\bibinfo{author}{\bibfnamefont{Y.}~\bibnamefont{Hara}} \bibnamefont{and}
  \bibinfo{author}{\bibfnamefont{S.}~\bibnamefont{Sakai}},
  \bibinfo{journal}{Phys. Lett. B} \textbf{\bibinfo{volume}{221}},
  \bibinfo{pages}{67} (\bibinfo{year}{1989}).

\bibitem[{\citenamefont{Robinett}(1991)}]{Robinett:1990qt}
\bibinfo{author}{\bibfnamefont{R.~W.} \bibnamefont{Robinett}},
  \bibinfo{journal}{Z. Phys. C} \textbf{\bibinfo{volume}{51}},
  \bibinfo{pages}{89} (\bibinfo{year}{1991}).

\bibitem[{\citenamefont{Jacobsen and Olsen}(1990)}]{Jacobsen:1990jp}
\bibinfo{author}{\bibfnamefont{T.}~\bibnamefont{Jacobsen}} \bibnamefont{and}
  \bibinfo{author}{\bibfnamefont{H.~A.} \bibnamefont{Olsen}},
  \bibinfo{journal}{Phys. Scripta} \textbf{\bibinfo{volume}{42}},
  \bibinfo{pages}{513} (\bibinfo{year}{1990}).

\bibitem[{\citenamefont{Groote et~al.}(1997)\citenamefont{Groote, Korner, and
  Leyva}}]{Groote:1997vg}
\bibinfo{author}{\bibfnamefont{S.}~\bibnamefont{Groote}},
  \bibinfo{author}{\bibfnamefont{J.~G.} \bibnamefont{Korner}},
  \bibnamefont{and} \bibinfo{author}{\bibfnamefont{J.~A.} \bibnamefont{Leyva}},
  \bibinfo{journal}{Phys. Rev. D} \textbf{\bibinfo{volume}{56}},
  \bibinfo{pages}{6031} (\bibinfo{year}{1997}), \eprint{hep-ph/9703416}.

\bibitem[{\citenamefont{Groote et~al.}(1999)\citenamefont{Groote, Korner, and
  Leyva}}]{Groote:1998qt}
\bibinfo{author}{\bibfnamefont{S.}~\bibnamefont{Groote}},
  \bibinfo{author}{\bibfnamefont{J.~G.} \bibnamefont{Korner}},
  \bibnamefont{and} \bibinfo{author}{\bibfnamefont{J.~A.} \bibnamefont{Leyva}},
  \bibinfo{journal}{Eur. Phys. J. C} \textbf{\bibinfo{volume}{7}},
  \bibinfo{pages}{49} (\bibinfo{year}{1999}), \eprint{hep-ph/9806464}.

\bibitem[{\citenamefont{Groote}(2002)}]{Groote:2002qc}
\bibinfo{author}{\bibfnamefont{S.}~\bibnamefont{Groote}}
  (\bibinfo{year}{2002}), \eprint{hep-ph/0212039}.

\bibitem[{\citenamefont{Berger et~al.}(2003)\citenamefont{Berger, Kucs, and
  Sterman}}]{Berger:2003iw}
\bibinfo{author}{\bibfnamefont{C.~F.} \bibnamefont{Berger}},
  \bibinfo{author}{\bibfnamefont{T.}~\bibnamefont{Kucs}}, \bibnamefont{and}
  \bibinfo{author}{\bibfnamefont{G.~F.} \bibnamefont{Sterman}},
  \bibinfo{journal}{Phys. Rev. D} \textbf{\bibinfo{volume}{68}},
  \bibinfo{pages}{014012} (\bibinfo{year}{2003}), \eprint{hep-ph/0303051}.

\bibitem[{\citenamefont{Almeida
  et~al.}(2009{\natexlab{a}})\citenamefont{Almeida, Lee, Perez, Sung, and
  Virzi}}]{Almeida:2008tp}
\bibinfo{author}{\bibfnamefont{L.~G.} \bibnamefont{Almeida}},
  \bibinfo{author}{\bibfnamefont{S.~J.} \bibnamefont{Lee}},
  \bibinfo{author}{\bibfnamefont{G.}~\bibnamefont{Perez}},
  \bibinfo{author}{\bibfnamefont{I.}~\bibnamefont{Sung}}, \bibnamefont{and}
  \bibinfo{author}{\bibfnamefont{J.}~\bibnamefont{Virzi}},
  \bibinfo{journal}{Phys. Rev. D} \textbf{\bibinfo{volume}{79}},
  \bibinfo{pages}{074012} (\bibinfo{year}{2009}{\natexlab{a}}),
  \eprint{0810.0934}.

\bibitem[{\citenamefont{Almeida
  et~al.}(2009{\natexlab{b}})\citenamefont{Almeida, Lee, Perez, Sterman, Sung,
  and Virzi}}]{Almeida:2008yp}
\bibinfo{author}{\bibfnamefont{L.~G.} \bibnamefont{Almeida}},
  \bibinfo{author}{\bibfnamefont{S.~J.} \bibnamefont{Lee}},
  \bibinfo{author}{\bibfnamefont{G.}~\bibnamefont{Perez}},
  \bibinfo{author}{\bibfnamefont{G.~F.} \bibnamefont{Sterman}},
  \bibinfo{author}{\bibfnamefont{I.}~\bibnamefont{Sung}}, \bibnamefont{and}
  \bibinfo{author}{\bibfnamefont{J.}~\bibnamefont{Virzi}},
  \bibinfo{journal}{Phys. Rev. D} \textbf{\bibinfo{volume}{79}},
  \bibinfo{pages}{074017} (\bibinfo{year}{2009}{\natexlab{b}}),
  \eprint{0807.0234}.

\bibitem[{\citenamefont{Nayak et~al.}(2005)\citenamefont{Nayak, Qiu, and
  Sterman}}]{Nayak:2005rt}
\bibinfo{author}{\bibfnamefont{G.~C.} \bibnamefont{Nayak}},
  \bibinfo{author}{\bibfnamefont{J.-W.} \bibnamefont{Qiu}}, \bibnamefont{and}
  \bibinfo{author}{\bibfnamefont{G.~F.} \bibnamefont{Sterman}},
  \bibinfo{journal}{Phys. Rev. D} \textbf{\bibinfo{volume}{72}},
  \bibinfo{pages}{114012} (\bibinfo{year}{2005}), \eprint{hep-ph/0509021}.

\bibitem[{\citenamefont{Collins}(2013)}]{Collins:2011zzd}
\bibinfo{author}{\bibfnamefont{J.}~\bibnamefont{Collins}},
  \emph{\bibinfo{title}{{Foundations of perturbative QCD}}},
  vol.~\bibinfo{volume}{32} (\bibinfo{publisher}{Cambridge University Press},
  \bibinfo{year}{2013}), ISBN \bibinfo{isbn}{978-1-107-64525-7,
  978-1-107-64525-7, 978-0-521-85533-4, 978-1-139-09782-6}.

\bibitem[{\citenamefont{Ellis et~al.}(2010)\citenamefont{Ellis, Vermilion,
  Walsh, Hornig, and Lee}}]{Ellis:2010rwa}
\bibinfo{author}{\bibfnamefont{S.~D.} \bibnamefont{Ellis}},
  \bibinfo{author}{\bibfnamefont{C.~K.} \bibnamefont{Vermilion}},
  \bibinfo{author}{\bibfnamefont{J.~R.} \bibnamefont{Walsh}},
  \bibinfo{author}{\bibfnamefont{A.}~\bibnamefont{Hornig}}, \bibnamefont{and}
  \bibinfo{author}{\bibfnamefont{C.}~\bibnamefont{Lee}},
  \bibinfo{journal}{JHEP} \textbf{\bibinfo{volume}{11}}, \bibinfo{pages}{101}
  (\bibinfo{year}{2010}), \eprint{1001.0014}.

\bibitem[{\citenamefont{Chen et~al.}(2021)\citenamefont{Chen, Moult, and
  Zhu}}]{Chen:2020adz}
\bibinfo{author}{\bibfnamefont{H.}~\bibnamefont{Chen}},
  \bibinfo{author}{\bibfnamefont{I.}~\bibnamefont{Moult}}, \bibnamefont{and}
  \bibinfo{author}{\bibfnamefont{H.~X.} \bibnamefont{Zhu}},
  \bibinfo{journal}{Phys. Rev. Lett.} \textbf{\bibinfo{volume}{126}},
  \bibinfo{pages}{112003} (\bibinfo{year}{2021}), \eprint{2011.02492}.

\bibitem[{\citenamefont{Chen et~al.}(2022)\citenamefont{Chen, Moult, and
  Zhu}}]{Chen:2021gdk}
\bibinfo{author}{\bibfnamefont{H.}~\bibnamefont{Chen}},
  \bibinfo{author}{\bibfnamefont{I.}~\bibnamefont{Moult}}, \bibnamefont{and}
  \bibinfo{author}{\bibfnamefont{H.~X.} \bibnamefont{Zhu}},
  \bibinfo{journal}{JHEP} \textbf{\bibinfo{volume}{08}}, \bibinfo{pages}{233}
  (\bibinfo{year}{2022}), \eprint{2104.00009}.

\bibitem[{\citenamefont{Larkoski}(2022)}]{Larkoski:2022lmv}
\bibinfo{author}{\bibfnamefont{A.~J.} \bibnamefont{Larkoski}},
  \bibinfo{journal}{Phys. Rev. D} \textbf{\bibinfo{volume}{105}},
  \bibinfo{pages}{096012} (\bibinfo{year}{2022}), \eprint{2201.03159}.

\bibitem[{\citenamefont{Gallicchio and Schwartz}(2011)}]{Gallicchio:2011xq}
\bibinfo{author}{\bibfnamefont{J.}~\bibnamefont{Gallicchio}} \bibnamefont{and}
  \bibinfo{author}{\bibfnamefont{M.~D.} \bibnamefont{Schwartz}},
  \bibinfo{journal}{Phys. Rev. Lett.} \textbf{\bibinfo{volume}{107}},
  \bibinfo{pages}{172001} (\bibinfo{year}{2011}), \eprint{1106.3076}.

\bibitem[{\citenamefont{Gallicchio and Schwartz}(2013)}]{Gallicchio:2012ez}
\bibinfo{author}{\bibfnamefont{J.}~\bibnamefont{Gallicchio}} \bibnamefont{and}
  \bibinfo{author}{\bibfnamefont{M.~D.} \bibnamefont{Schwartz}},
  \bibinfo{journal}{JHEP} \textbf{\bibinfo{volume}{04}}, \bibinfo{pages}{090}
  (\bibinfo{year}{2013}), \eprint{1211.7038}.

\bibitem[{\citenamefont{Ferreira~de Lima et~al.}(2017)\citenamefont{Ferreira~de
  Lima, Petrov, Soper, and Spannowsky}}]{FerreiradeLima:2016gcz}
\bibinfo{author}{\bibfnamefont{D.}~\bibnamefont{Ferreira~de Lima}},
  \bibinfo{author}{\bibfnamefont{P.}~\bibnamefont{Petrov}},
  \bibinfo{author}{\bibfnamefont{D.}~\bibnamefont{Soper}}, \bibnamefont{and}
  \bibinfo{author}{\bibfnamefont{M.}~\bibnamefont{Spannowsky}},
  \bibinfo{journal}{Phys. Rev. D} \textbf{\bibinfo{volume}{95}},
  \bibinfo{pages}{034001} (\bibinfo{year}{2017}), \eprint{1607.06031}.

\bibitem[{\citenamefont{Frye et~al.}(2017)\citenamefont{Frye, Larkoski, Thaler,
  and Zhou}}]{Frye:2017yrw}
\bibinfo{author}{\bibfnamefont{C.}~\bibnamefont{Frye}},
  \bibinfo{author}{\bibfnamefont{A.~J.} \bibnamefont{Larkoski}},
  \bibinfo{author}{\bibfnamefont{J.}~\bibnamefont{Thaler}}, \bibnamefont{and}
  \bibinfo{author}{\bibfnamefont{K.}~\bibnamefont{Zhou}},
  \bibinfo{journal}{JHEP} \textbf{\bibinfo{volume}{09}}, \bibinfo{pages}{083}
  (\bibinfo{year}{2017}), \eprint{1704.06266}.

\bibitem[{\citenamefont{Banfi et~al.}(2006)\citenamefont{Banfi, Salam, and
  Zanderighi}}]{Banfi:2006hf}
\bibinfo{author}{\bibfnamefont{A.}~\bibnamefont{Banfi}},
  \bibinfo{author}{\bibfnamefont{G.~P.} \bibnamefont{Salam}}, \bibnamefont{and}
  \bibinfo{author}{\bibfnamefont{G.}~\bibnamefont{Zanderighi}},
  \bibinfo{journal}{Eur. Phys. J. C} \textbf{\bibinfo{volume}{47}},
  \bibinfo{pages}{113} (\bibinfo{year}{2006}), \eprint{hep-ph/0601139}.

\bibitem[{\citenamefont{Gras et~al.}(2017)\citenamefont{Gras, H\"oche, Kar,
  Larkoski, L\"onnblad, Pl\"atzer, Si\'odmok, Skands, Soyez, and
  Thaler}}]{Gras:2017jty}
\bibinfo{author}{\bibfnamefont{P.}~\bibnamefont{Gras}},
  \bibinfo{author}{\bibfnamefont{S.}~\bibnamefont{H\"oche}},
  \bibinfo{author}{\bibfnamefont{D.}~\bibnamefont{Kar}},
  \bibinfo{author}{\bibfnamefont{A.}~\bibnamefont{Larkoski}},
  \bibinfo{author}{\bibfnamefont{L.}~\bibnamefont{L\"onnblad}},
  \bibinfo{author}{\bibfnamefont{S.}~\bibnamefont{Pl\"atzer}},
  \bibinfo{author}{\bibfnamefont{A.}~\bibnamefont{Si\'odmok}},
  \bibinfo{author}{\bibfnamefont{P.}~\bibnamefont{Skands}},
  \bibinfo{author}{\bibfnamefont{G.}~\bibnamefont{Soyez}}, \bibnamefont{and}
  \bibinfo{author}{\bibfnamefont{J.}~\bibnamefont{Thaler}},
  \bibinfo{journal}{JHEP} \textbf{\bibinfo{volume}{07}}, \bibinfo{pages}{091}
  (\bibinfo{year}{2017}), \eprint{1704.03878}.

\bibitem[{\citenamefont{Metodiev and Thaler}(2018)}]{Metodiev:2018ftz}
\bibinfo{author}{\bibfnamefont{E.~M.} \bibnamefont{Metodiev}} \bibnamefont{and}
  \bibinfo{author}{\bibfnamefont{J.}~\bibnamefont{Thaler}},
  \bibinfo{journal}{Phys. Rev. Lett.} \textbf{\bibinfo{volume}{120}},
  \bibinfo{pages}{241602} (\bibinfo{year}{2018}), \eprint{1802.00008}.

\bibitem[{\citenamefont{Larkoski et~al.}(2014)\citenamefont{Larkoski, Thaler,
  and Waalewijn}}]{Larkoski:2014pca}
\bibinfo{author}{\bibfnamefont{A.~J.} \bibnamefont{Larkoski}},
  \bibinfo{author}{\bibfnamefont{J.}~\bibnamefont{Thaler}}, \bibnamefont{and}
  \bibinfo{author}{\bibfnamefont{W.~J.} \bibnamefont{Waalewijn}},
  \bibinfo{journal}{JHEP} \textbf{\bibinfo{volume}{11}}, \bibinfo{pages}{129}
  (\bibinfo{year}{2014}), \eprint{1408.3122}.

\bibitem[{\citenamefont{Bhattacherjee et~al.}(2015)\citenamefont{Bhattacherjee,
  Mukhopadhyay, Nojiri, Sakaki, and Webber}}]{Bhattacherjee:2015psa}
\bibinfo{author}{\bibfnamefont{B.}~\bibnamefont{Bhattacherjee}},
  \bibinfo{author}{\bibfnamefont{S.}~\bibnamefont{Mukhopadhyay}},
  \bibinfo{author}{\bibfnamefont{M.~M.} \bibnamefont{Nojiri}},
  \bibinfo{author}{\bibfnamefont{Y.}~\bibnamefont{Sakaki}}, \bibnamefont{and}
  \bibinfo{author}{\bibfnamefont{B.~R.} \bibnamefont{Webber}},
  \bibinfo{journal}{JHEP} \textbf{\bibinfo{volume}{04}}, \bibinfo{pages}{131}
  (\bibinfo{year}{2015}), \eprint{1501.04794}.

\bibitem[{\citenamefont{Kasieczka et~al.}(2019)\citenamefont{Kasieczka, Kiefer,
  Plehn, and Thompson}}]{Kasieczka:2018lwf}
\bibinfo{author}{\bibfnamefont{G.}~\bibnamefont{Kasieczka}},
  \bibinfo{author}{\bibfnamefont{N.}~\bibnamefont{Kiefer}},
  \bibinfo{author}{\bibfnamefont{T.}~\bibnamefont{Plehn}}, \bibnamefont{and}
  \bibinfo{author}{\bibfnamefont{J.~M.} \bibnamefont{Thompson}},
  \bibinfo{journal}{SciPost Phys.} \textbf{\bibinfo{volume}{6}},
  \bibinfo{pages}{069} (\bibinfo{year}{2019}), \eprint{1812.09223}.

\bibitem[{\citenamefont{Larkoski and Metodiev}(2019)}]{Larkoski:2019nwj}
\bibinfo{author}{\bibfnamefont{A.~J.} \bibnamefont{Larkoski}} \bibnamefont{and}
  \bibinfo{author}{\bibfnamefont{E.~M.} \bibnamefont{Metodiev}},
  \bibinfo{journal}{JHEP} \textbf{\bibinfo{volume}{10}}, \bibinfo{pages}{014}
  (\bibinfo{year}{2019}), \eprint{1906.01639}.

\bibitem[{\citenamefont{Bright-Thonney
  et~al.}(2022)\citenamefont{Bright-Thonney, Moult, Nachman, and
  Prestel}}]{Bright-Thonney:2022xkx}
\bibinfo{author}{\bibfnamefont{S.}~\bibnamefont{Bright-Thonney}},
  \bibinfo{author}{\bibfnamefont{I.}~\bibnamefont{Moult}},
  \bibinfo{author}{\bibfnamefont{B.}~\bibnamefont{Nachman}}, \bibnamefont{and}
  \bibinfo{author}{\bibfnamefont{S.}~\bibnamefont{Prestel}}
  (\bibinfo{year}{2022}), \eprint{2207.12411}.

\bibitem[{\citenamefont{Collins et~al.}(1994)\citenamefont{Collins, Heppelmann,
  and Ladinsky}}]{Collins:1993kq}
\bibinfo{author}{\bibfnamefont{J.~C.} \bibnamefont{Collins}},
  \bibinfo{author}{\bibfnamefont{S.~F.} \bibnamefont{Heppelmann}},
  \bibnamefont{and} \bibinfo{author}{\bibfnamefont{G.~A.}
  \bibnamefont{Ladinsky}}, \bibinfo{journal}{Nucl. Phys. B}
  \textbf{\bibinfo{volume}{420}}, \bibinfo{pages}{565} (\bibinfo{year}{1994}),
  \eprint{hep-ph/9305309}.

\bibitem[{\citenamefont{Kang et~al.}(2020)\citenamefont{Kang, Lee, and
  Zhao}}]{Kang:2020xyq}
\bibinfo{author}{\bibfnamefont{Z.-B.} \bibnamefont{Kang}},
  \bibinfo{author}{\bibfnamefont{K.}~\bibnamefont{Lee}}, \bibnamefont{and}
  \bibinfo{author}{\bibfnamefont{F.}~\bibnamefont{Zhao}},
  \bibinfo{journal}{Phys. Lett. B} \textbf{\bibinfo{volume}{809}},
  \bibinfo{pages}{135756} (\bibinfo{year}{2020}), \eprint{2005.02398}.

\bibitem[{\citenamefont{Yu and Yuan}(2022)}]{Yu:2021zmw}
\bibinfo{author}{\bibfnamefont{Z.}~\bibnamefont{Yu}} \bibnamefont{and}
  \bibinfo{author}{\bibfnamefont{C.~P.} \bibnamefont{Yuan}},
  \bibinfo{journal}{Phys. Rev. Lett.} \textbf{\bibinfo{volume}{129}},
  \bibinfo{pages}{112001} (\bibinfo{year}{2022}), \eprint{2110.11539}.

\bibitem[{\citenamefont{Hou et~al.}(2021)}]{Hou:2019efy}
\bibinfo{author}{\bibfnamefont{T.-J.} \bibnamefont{Hou}} \bibnamefont{et~al.},
  \bibinfo{journal}{Phys. Rev. D} \textbf{\bibinfo{volume}{103}},
  \bibinfo{pages}{014013} (\bibinfo{year}{2021}), \eprint{1912.10053}.

\bibitem[{\citenamefont{Alwall et~al.}(2014)\citenamefont{Alwall, Frederix,
  Frixione, Hirschi, Maltoni, Mattelaer, Shao, Stelzer, Torrielli, and
  Zaro}}]{Alwall:2014hca}
\bibinfo{author}{\bibfnamefont{J.}~\bibnamefont{Alwall}},
  \bibinfo{author}{\bibfnamefont{R.}~\bibnamefont{Frederix}},
  \bibinfo{author}{\bibfnamefont{S.}~\bibnamefont{Frixione}},
  \bibinfo{author}{\bibfnamefont{V.}~\bibnamefont{Hirschi}},
  \bibinfo{author}{\bibfnamefont{F.}~\bibnamefont{Maltoni}},
  \bibinfo{author}{\bibfnamefont{O.}~\bibnamefont{Mattelaer}},
  \bibinfo{author}{\bibfnamefont{H.~S.} \bibnamefont{Shao}},
  \bibinfo{author}{\bibfnamefont{T.}~\bibnamefont{Stelzer}},
  \bibinfo{author}{\bibfnamefont{P.}~\bibnamefont{Torrielli}},
  \bibnamefont{and} \bibinfo{author}{\bibfnamefont{M.}~\bibnamefont{Zaro}},
  \bibinfo{journal}{JHEP} \textbf{\bibinfo{volume}{07}}, \bibinfo{pages}{079}
  (\bibinfo{year}{2014}), \eprint{1405.0301}.

\bibitem[{CMS(2016)}]{CMS-PAS-BTV-15-002}
\bibinfo{type}{Tech. Rep.}, \bibinfo{institution}{CERN},
  \bibinfo{address}{Geneva} (\bibinfo{year}{2016}),
  \urlprefix\url{http://cds.cern.ch/record/2195743}.

\bibitem[{ATL(2016)}]{ATLAS-CONF-2016-002}
\bibinfo{type}{Tech. Rep.}, \bibinfo{institution}{CERN},
  \bibinfo{address}{Geneva} (\bibinfo{year}{2016}).

\bibitem[{\citenamefont{Sirunyan et~al.}(2018)}]{CMS:2017wtu}
\bibinfo{author}{\bibfnamefont{A.~M.} \bibnamefont{Sirunyan}}
  \bibnamefont{et~al.} (\bibinfo{collaboration}{CMS}), \bibinfo{journal}{JINST}
  \textbf{\bibinfo{volume}{13}}, \bibinfo{pages}{P05011}
  (\bibinfo{year}{2018}), \eprint{1712.07158}.

\bibitem[{\citenamefont{Aaboud et~al.}(2018{\natexlab{a}})}]{ATLAS:2018sgt}
\bibinfo{author}{\bibfnamefont{M.}~\bibnamefont{Aaboud}} \bibnamefont{et~al.}
  (\bibinfo{collaboration}{ATLAS}), \bibinfo{journal}{JHEP}
  \textbf{\bibinfo{volume}{08}}, \bibinfo{pages}{089}
  (\bibinfo{year}{2018}{\natexlab{a}}), \eprint{1805.01845}.

\bibitem[{\citenamefont{Aaboud et~al.}(2018{\natexlab{b}})}]{ATLAS:2018mgv}
\bibinfo{author}{\bibfnamefont{M.}~\bibnamefont{Aaboud}} \bibnamefont{et~al.}
  (\bibinfo{collaboration}{ATLAS}), \bibinfo{journal}{Phys. Rev. Lett.}
  \textbf{\bibinfo{volume}{120}}, \bibinfo{pages}{211802}
  (\bibinfo{year}{2018}{\natexlab{b}}), \eprint{1802.04329}.

\bibitem[{\citenamefont{Aad et~al.}(2019{\natexlab{a}})}]{ATLAS:2019bwq}
\bibinfo{author}{\bibfnamefont{G.}~\bibnamefont{Aad}} \bibnamefont{et~al.}
  (\bibinfo{collaboration}{ATLAS}), \bibinfo{journal}{Eur. Phys. J. C}
  \textbf{\bibinfo{volume}{79}}, \bibinfo{pages}{970}
  (\bibinfo{year}{2019}{\natexlab{a}}), \eprint{1907.05120}.

\bibitem[{\citenamefont{Aad et~al.}(2019{\natexlab{b}})}]{ATLAS:2019lwq}
\bibinfo{author}{\bibfnamefont{G.}~\bibnamefont{Aad}} \bibnamefont{et~al.}
  (\bibinfo{collaboration}{ATLAS}), \bibinfo{journal}{Eur. Phys. J. C}
  \textbf{\bibinfo{volume}{79}}, \bibinfo{pages}{836}
  (\bibinfo{year}{2019}{\natexlab{b}}), \eprint{1906.11005}.

\bibitem[{\citenamefont{Tumasyan et~al.}(2022)}]{CMS:2021scf}
\bibinfo{author}{\bibfnamefont{A.}~\bibnamefont{Tumasyan}} \bibnamefont{et~al.}
  (\bibinfo{collaboration}{CMS}), \bibinfo{journal}{JINST}
  \textbf{\bibinfo{volume}{17}}, \bibinfo{pages}{P03014}
  (\bibinfo{year}{2022}), \eprint{2111.03027}.

\bibitem[{\citenamefont{Aad et~al.}(2022)}]{ATLAS:2021cxe}
\bibinfo{author}{\bibfnamefont{G.}~\bibnamefont{Aad}} \bibnamefont{et~al.}
  (\bibinfo{collaboration}{ATLAS}), \bibinfo{journal}{Eur. Phys. J. C}
  \textbf{\bibinfo{volume}{82}}, \bibinfo{pages}{95} (\bibinfo{year}{2022}),
  \eprint{2109.10627}.

\bibitem[{ATL(2022{\natexlab{a}})}]{ATL-PHYS-PUB-2022-027}
\bibinfo{type}{Tech. Rep.}, \bibinfo{institution}{CERN},
  \bibinfo{address}{Geneva} (\bibinfo{year}{2022}{\natexlab{a}}),
  \urlprefix\url{https://cds.cern.ch/record/2811135}.

\bibitem[{ATL(2022{\natexlab{b}})}]{ATL-PHYS-PUB-2022-010}
\bibinfo{type}{Tech. Rep.}, \bibinfo{institution}{CERN},
  \bibinfo{address}{Geneva} (\bibinfo{year}{2022}{\natexlab{b}}),
  \urlprefix\url{https://cds.cern.ch/record/2804062}.

\bibitem[{\citenamefont{Sj\"ostrand et~al.}(2015)\citenamefont{Sj\"ostrand,
  Ask, Christiansen, Corke, Desai, Ilten, Mrenna, Prestel, Rasmussen, and
  Skands}}]{Sjostrand:2014zea}
\bibinfo{author}{\bibfnamefont{T.}~\bibnamefont{Sj\"ostrand}},
  \bibinfo{author}{\bibfnamefont{S.}~\bibnamefont{Ask}},
  \bibinfo{author}{\bibfnamefont{J.~R.} \bibnamefont{Christiansen}},
  \bibinfo{author}{\bibfnamefont{R.}~\bibnamefont{Corke}},
  \bibinfo{author}{\bibfnamefont{N.}~\bibnamefont{Desai}},
  \bibinfo{author}{\bibfnamefont{P.}~\bibnamefont{Ilten}},
  \bibinfo{author}{\bibfnamefont{S.}~\bibnamefont{Mrenna}},
  \bibinfo{author}{\bibfnamefont{S.}~\bibnamefont{Prestel}},
  \bibinfo{author}{\bibfnamefont{C.~O.} \bibnamefont{Rasmussen}},
  \bibnamefont{and} \bibinfo{author}{\bibfnamefont{P.~Z.}
  \bibnamefont{Skands}}, \bibinfo{journal}{Comput. Phys. Commun.}
  \textbf{\bibinfo{volume}{191}}, \bibinfo{pages}{159} (\bibinfo{year}{2015}),
  \eprint{1410.3012}.

\bibitem[{\citenamefont{Aaboud et~al.}(2019)}]{ATLAS:2018zhf}
\bibinfo{author}{\bibfnamefont{M.}~\bibnamefont{Aaboud}} \bibnamefont{et~al.}
  (\bibinfo{collaboration}{ATLAS}), \bibinfo{journal}{Phys. Rev. D}
  \textbf{\bibinfo{volume}{99}}, \bibinfo{pages}{052004}
  (\bibinfo{year}{2019}), \eprint{1812.09283}.

\end{thebibliography}

\end{document}